\documentclass[]{article}
\usepackage{lmodern}
\usepackage{amssymb,amsmath}
\usepackage{ifxetex,ifluatex}
\usepackage{fixltx2e} 
\ifnum 0\ifxetex 1\fi\ifluatex 1\fi=0 
  \usepackage[T1]{fontenc}
  \usepackage[utf8]{inputenc}
\else 
  \ifxetex
    \usepackage{mathspec}
  \else
    \usepackage{fontspec}
  \fi
  \defaultfontfeatures{Ligatures=TeX,Scale=MatchLowercase}
\fi
\IfFileExists{upquote.sty}{\usepackage{upquote}}{}
\IfFileExists{microtype.sty}{%
\usepackage{microtype}
\UseMicrotypeSet[protrusion]{basicmath} 
}{}
\usepackage[margin=4cm]{geometry}
\usepackage{hyperref}
\hypersetup{unicode=true,
            pdftitle={Exploratory Mediation Analysis with Many Potential Mediators},
            pdfauthor={Erik-Jan van Kesteren and Daniel Oberski},
            pdfborder={0 0 0},
            breaklinks=true}
\usepackage{longtable,booktabs}
\usepackage{graphicx,grffile}
\makeatletter
\def\maxwidth{\ifdim\Gin@nat@width>\linewidth\linewidth\else\Gin@nat@width\fi}
\def\maxheight{\ifdim\Gin@nat@height>\textheight\textheight\else\Gin@nat@height\fi}
\makeatother
\setkeys{Gin}{width=\maxwidth,height=\maxheight,keepaspectratio}
\IfFileExists{parskip.sty}{%
\usepackage{parskip}
}{
\setlength{\parindent}{0pt}
\setlength{\parskip}{6pt plus 2pt minus 1pt}
}
\ifx\paragraph\undefined\else
\let\oldparagraph\paragraph
\renewcommand{\paragraph}[1]{\oldparagraph{#1}\mbox{}}
\fi
\ifx\subparagraph\undefined\else
\let\oldsubparagraph\subparagraph
\renewcommand{\subparagraph}[1]{\oldsubparagraph{#1}\mbox{}}
\fi

\let\rmarkdownfootnote\footnote%
\def\footnote{\protect\rmarkdownfootnote}

\usepackage{titling}

  \title{Exploratory Mediation Analysis with Many Potential Mediators}
    \pretitle{\vspace{\droptitle}\centering\huge}
  \posttitle{\par}
    \author{Erik-Jan van Kesteren and Daniel Oberski}
    \preauthor{\centering\large\emph}
  \postauthor{\par}
      \predate{\centering\large\emph}
  \postdate{\par}
    \date{Utrecht University, Department of Methodology and Statistics}

\usepackage{booktabs}
\usepackage{longtable}
\usepackage{array}
\usepackage{multirow}
\usepackage{wrapfig}
\usepackage{float}
\usepackage{colortbl}
\usepackage{pdflscape}
\usepackage{tabu}
\usepackage{threeparttable}
\usepackage{threeparttablex}
\usepackage[normalem]{ulem}
\usepackage{makecell}
\usepackage{xcolor}

\usepackage{algorithm}
\usepackage[noend]{algpseudocode}
\usepackage{float}
\usepackage{caption}
\begin{document}
\maketitle
\begin{abstract}
Social and behavioral scientists are increasingly employing technologies such as fMRI, smartphones, and gene sequencing, which yield `high-dimensional' datasets with more columns than rows. There is increasing interest, but little substantive theory, in the role the variables in these data play in known processes.

This necessitates exploratory mediation analysis, for which structural equation modeling is the benchmark method. However, this method cannot perform mediation analysis with more variables than observations. One option is to run a series of univariate mediation models, which incorrectly assumes independence of the mediators. Another option is regularization, but the available implementations may lead to high false positive rates.

In this paper, we develop a hybrid approach which uses components of both filter and regularization: the `Coordinate-wise Mediation Filter'. It performs filtering conditional on the other selected mediators. We show through simulation that it improves performance over existing methods. Finally, we provide an empirical example, showing how our method may be used for epigenetic research.
\end{abstract}

\hypertarget{introduction}{%
\section{Introduction}\label{introduction}}

Social and behavioral scientists are increasingly employing technologies such as fMRI, smartphones, and gene sequencing, which yield `high-dimensional' datasets with more variables than observations. These high-dimensional data are often intended to answer questions such as \emph{``which areas of our brain are relevant for pain perception?''} (Atlas, Lindquist, Bolger, \& Wager, 2014) and \emph{``which genes mediate the effect of trauma on stress reactivity?''} (Houtepen et al., 2016). These are questions regarding exploratory mediation analysis (EMA).

\begin{figure}[H]

{\centering \includegraphics[width=0.6\linewidth]{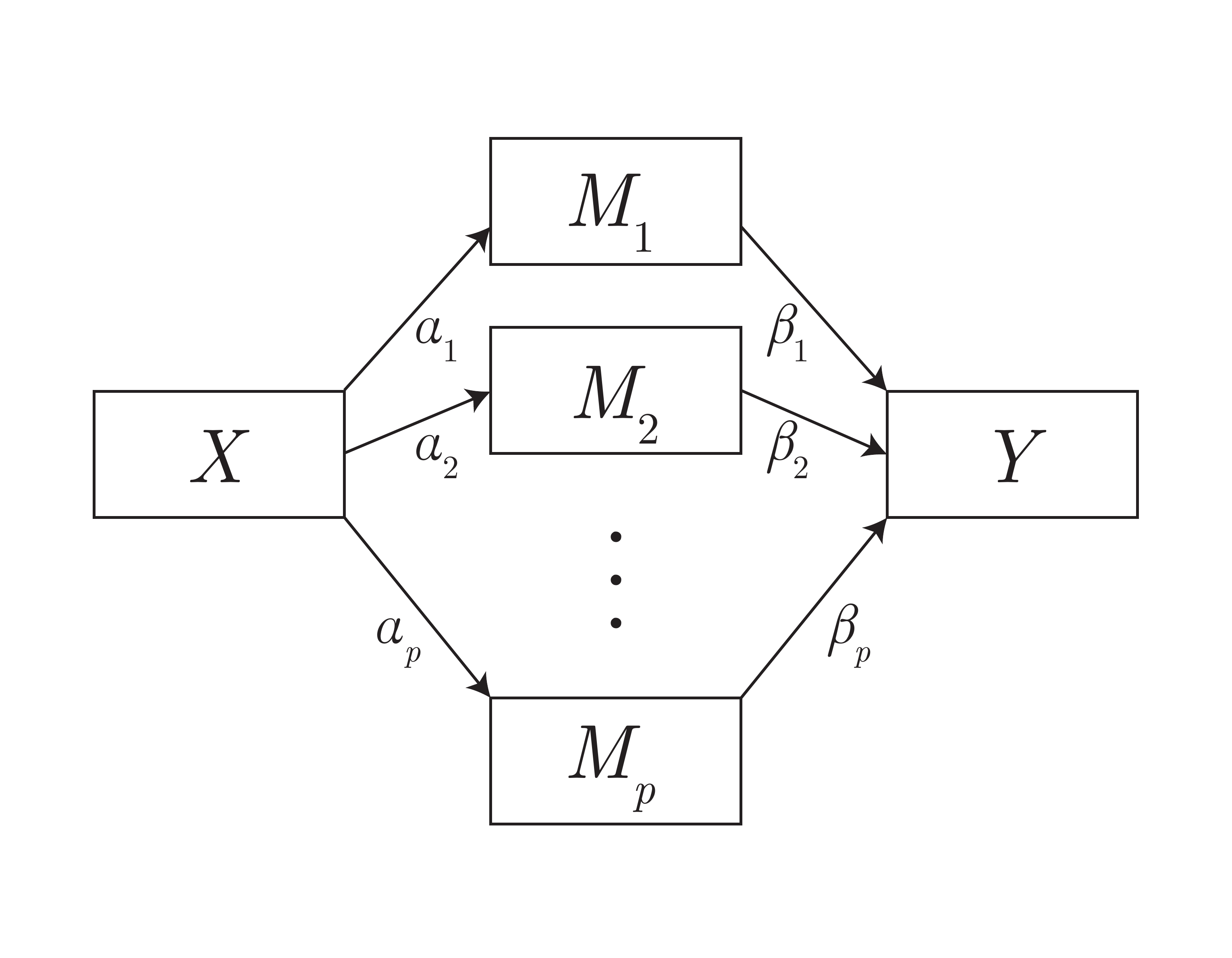} 

}

\caption{Exploratory mediation analysis with a set of \(p\) potential mediators \(\boldsymbol{M}\). For clarity, we omitted the \(P(P+1)/2\) parameters belonging to the residuals of \(\boldsymbol{M}\) and their covariances, as well as the residual variance of \(Y\).}\label{fig:ema}
\end{figure}

Structural equation modeling (SEM) is the preferred method for mediation analysis with multiple mediators (Preacher \& Hayes, 2008; Vanderweele \& Vansteelandt, 2014). With this method, it is possible to determine to what extent specific \(M\) variables mediate the \(X \to Y\) effect conditional on the presence of other mediators in the model. However, this method fails when the data is \emph{high-dimensional}, when the variables under investigation outnumber the samples \(N\). In this situation, the observed covariance matrix is rank-deficient, leading to linear dependence in the observed moments and, for the full mediation model, nonconvergence.

Several alternative methods for EMA have been proposed to deal with this issue. One option mentioned by Preacher \& Hayes (2008) is to select relevant mediators from a series of univariate \(X\to M \to Y\) mediation models (e.g. Boca, Sinha, Cross, Moore, \& Sampson, 2014; Liu et al., 2013). We call this the ``filter'' method, following the taxonomy of Guyon \& Elisseeff (2003). Its main advantages are that it is simple to explain and run, requiring only \(P\) univariate path models. On the other hand, the filter method introduces bias through model misspecification: it takes into account only the \emph{marginal} relationships of \(M\) with \(X\) and \(Y\). A pitfall of this is that a variable useless by itself can be useful together with others (Guyon \& Elisseeff, 2003). In other words, a certain mediator may be marginally irrelevant, but relevant conditional on another set of mediators.

Recently, another multivariate method was introduced by Serang, Jacobucci, Brimhall, \& Grimm (2017). Their proposal was to perform EMA through regularized estimation of the full structural equation model: ``XMed''. This method automatically shrinks small regression paths to 0, leading to a selection of potential mediators: mediators are variables for which both the \(X \to M\) path and the \(M\to Y\) path are nonzero after regularization. With this method, it is possible to detect mediators which are only relevant conditionally, while regularization resolves the identification issues of default SEM (Hastie, Tibshirani, \& Wainwright, 2015). The disadvantage is that this method finds paths with a large effect rather than the desired subset of mediators: the regularization in XMed shrinks small \(\beta\) paths to 0, irrespective of the value of their associated \(\alpha\) paths -- shrinkage is performed on all paths equally. This leads to inflated false positive rates as reported by Serang et al. (2017) and Jacobucci, Brandmaier, \& Kievit (2018). In summary, regularization methods do perform conditional estimation, but they select paths rather than mediators.

In this paper, we propose a hybrid approach to EMA which we call the ``Coordinate-wise Mediation Filter'' (CMF). This method combines advantages from both the filter and regularization methods: (a) it converges in case of high-dimensional data, (b) it takes into account mediator correlations, leading to conditional selection of mediators, and (c) it selects based on mediation, not paths. CMF performs univariate filtering \emph{conditional} on the other selected mediators by using an algorithm from regularized regression: cyclical coordinate descent on residuals (Breheny \& Huang, 2011; Friedman, Hastie, \& Tibshirani, 2009).

The remainder of the article is structured as follows: first, we provide relevant background on exploratory mediation analysis. Then, we outline the Coordinate-wise Mediation Filter as a hybrid method for mediator subset selection. Following this, we show through simulation where each of the discussed methods performs as well as SEM. In addition, we assess the performance of CMF relative to the other available methods in a high-dimensional simulation. Lastly, the CMF procedure is illustrated by applying it to the epigenetic process of trauma and stress reactivity.

\hypertarget{exploratory-mediation-analysis}{%
\subsection{Exploratory mediation analysis}\label{exploratory-mediation-analysis}}

The fundamental goal of mediation analysis is to determine the process by which a variable \(X\) influences another variable \(Y\) (MacKinnon, Lockwood, \& Williams, 2004). Exploratory mediation analysis (EMA) in particular is used to explore a dataset for potential mediating variables (MacKinnon, 2008). In other words, EMA pertains to determining among multiple potential mediators which subset is most relevant. Through EMA, researchers can build theory and select variables of interest for further research into the process under investigation.

An example application of EMA is the research by Ammerman et al. (2018), who investigated how childhood maltreatment leads to suicidal behaviour. They defined 46 potential mediators, including psychological counseling, closeness to parents, and self-esteem. The authors did not test a fully specified mediation model about the precise relations of each of these variables to childhood maltreatment and suicidal behaviour. Instead, this study was exploratory, identifying which variables were the most relevant targets for future research. Indeed, the authors conclude that the study ``highlights factors that may be potential targets for risk assessment and for treatment among adolescents with a history of childhood maltreatment''.

\hypertarget{univariate-mediation-analysis-and-the-filter-method}{%
\subsubsection{Univariate mediation analysis and the filter method}\label{univariate-mediation-analysis-and-the-filter-method}}

\begin{figure}[H]

{\centering \includegraphics[width=0.6\linewidth]{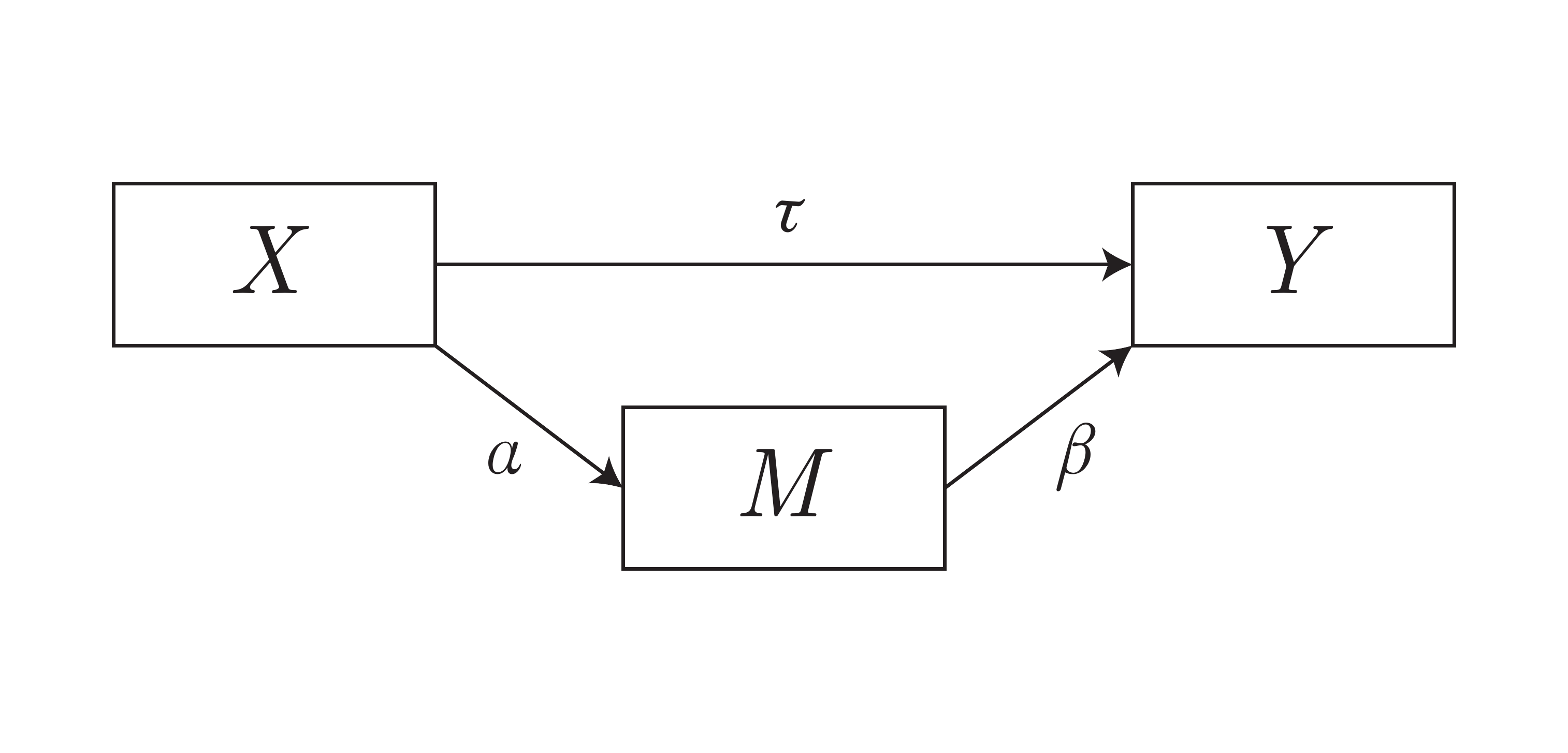} 

}

\caption{Graphical representation of the system of Equation \eqref{eq:syslin1}. For clarity, the residuals are not shown.}\label{fig:sys}
\end{figure}

A common framework for \emph{univariate} mediation analysis is a system of regression equations (Equation \eqref{eq:syslin1}; MacKinnon et al., 2004). The system is displayed graphically in Figure \ref{fig:sys}. In the present paper, we consider only the case where the data from \(X\), \(M\), and \(Y\) are continuous and their relations are linear. For nonlinear discrete extensions to mediation analysis, see Hayes \& Preacher (2010) and Hayes \& Preacher (2014), respectively. For further details, refer to the reviews by MacKinnon, Fairchild, \& Fritz (2007) and Preacher (2015).
\begin{align}
M &= \mu_M + \alpha X + e_M \nonumber \\
Y &= \mu_Y + \tau X + \beta M + e_Y 
\label{eq:syslin1}
\end{align}

Under the standard assumptions of linear SEM, the parameter estimates of this system may be used to determine whether \(M\) is a mediator --- a dichotomous decision. There are several ways to make this decision, usually based on a quantity of interest \(q\) and a measure of uncertainty (MacKinnon, Lockwood, Hoffman, West, \& Sheets, 2002). For example, \(q\) may represent the size of the indirect effect through the product of its coefficients \(q_{\text{prod}} = \alpha\beta\), and uncertainty measures for \(q_{\text{prod}}\) can be obtained using asymptotic standard error methods (e.g., Olkin \& Finn, 1995; Sobel, 1986) or bootstrapping (Preacher \& Hayes, 2008).

Combining the quantity of interest \(q\) with an uncertainty estimate and a specified alpha level yields a dichotomous decision criterion based on a \(p\)-value. We call this a \emph{univariate decision function} \(\mathcal{D}\): a function that maps the data of \(X\), \(M\), and \(Y\) to a binary decision of whether \(M\) should be considered a mediator (1) or not (0).
\[\mathcal{D}: (\boldsymbol{x}, \boldsymbol{m}, \boldsymbol{y}) \mapsto \{0, 1\}\]
Note that any function that follows this specification can be considered a decision function, regardless of complexity. An example of higher complexity decision functions is given by VanderWeele (2015, p. 46), who states exposure-outcome confounding should by default be controlled for when testing for mediation. The decision function encodes the researcher's definition of mediation: a product of coefficients decision function with a \(p\)-value cutoff of 0.1 will lead to different results than an exposure-outcome controlled decision function with a stricter cutoff.

This decision function framework thus provides a convenient abstraction, highlighting a key advantage for mediation analysis methods: if the choice of decision functions is flexible, a method is adaptable to the specific needs of a researcher. If researchers want to follow the recommendation of VanderWeele (2015), they can do so by adding an \(XM\) interaction term into the decision function.

While these decision functions are univariate, EMA is an inherently multivariate procedure, requiring analysis of multiple indirect effects. To perform EMA, a researcher can apply their chosen decision function to each mediator separately, through \(P\) different mediation models as in Figure \ref{fig:sys}. This ``filter method'' will result in a subset of relevant mediators. However, the implicit assumption is that the \(M\) variables are independent of one another. In other words, the selected subset will not include mediators that are relevant only conditionally on another mediator.

\hypertarget{multivariate-mediation-analysis-and-xmed}{%
\subsubsection{Multivariate mediation analysis and XMed}\label{multivariate-mediation-analysis-and-xmed}}

To make mediation decisions multivariately, Preacher \& Hayes (2008) recommend the SEM approach. In this approach, the quantities of interest \(q_1, ..., q_P\) and their uncertainty are estimated directly from a multiple mediation model as in Figure \ref{fig:ema}. A decision can then be made for each individual \(M_p\) based on its multivariately estimated quantity \(q_p\). Unlike the filter method, this approach estimates \(q_p\) conditional on the other \(P-1\) quantities, so that marginally irrelevant true mediators may still be detected.

However, the SEM approach is unavailable in the case of high-dimensional data because SEM parameters are estimated from observed covariances. High dimensional data \((P > N)\) leads to a \(P\times P\) observed covariance matrix of at most rank \(N\), meaning a linear dependence exists among elements. If dependent elements are mapped to separate parameters in the SEM model, an infinite number of solutions exist for the same log-likelihood, so there is no maximum likelihood solution. This is the case in the full mediation model. As an alternative intuitive explanation, it is possible to view the \(M\to Y\) part of the mediation model as a high-dimensional multiple regression, where ordinary least squares (OLS) estimates are unavailable because the covariance matrix cannot be inverted (Hastie et al., 2015).

XMed is an adjustment to the SEM method that not only allows for high-dimensional data, but it also automatically selects a subset of mediators without an explicit decision function. The estimation method for XMed is RegSEM (Jacobucci, Grimm, \& McArdle, 2016), which applies regularization to a chosen subset of model parameters in a structural equation model. This shrinkage is determined by the hyperparameter \(\lambda\) along with the penalization function \(P(\cdot)\) in the objective function of RegSEM:
\[F_{regsem} = F_{ML} + \lambda P(\cdot)\] where \(\cdot\) is a vector of parameters.

In XMed specifically, shrinkage is applied to the vectors of \(\alpha\) (\(x \to M\)) and \(\beta\) (\(M \to y\)) parameters. Subset selection of the mediators occurs through the chosen regularization method; the penalty function \(P(\cdot)\) is the LASSO penalty, the \(\ell_1\) norm of the chosen parameter vector: \(P(\boldsymbol{\cdot}) = \| \cdot \|_1\). Depending on the value of \(\lambda\), The LASSO penalty shrinks the smallest of the chosen parameters to 0 during estimation. This immediately forms the decision rule: for potential mediator \(\boldsymbol{M}_p\), if \(\alpha_p\) or \(\beta_p\) equals 0, then the estimated indirect effect \(\alpha\beta_p\) is 0, thus \(\boldsymbol{M}_p\) is not considered to be a true mediator.

A well-known algorithm for computing the LASSO solution, which can also be applied in SEM, is coordinate-wise conditioning or coordinate descent: the conditional solution is well-known and easy to find, in SEM the maximum likelihood estimates, and the penalized solution is found by cyclically updating and soft-thresholding the conditional solution for each parameter in turn, until convergence (Hastie et al., 2015).

A sequential combination of the ideas of filtering and regularization was proposed by Zhang et al. (2016) in a three-step approach called HIMA. First, in the screening step the authors marginally filter irrelevant potential mediators based on the \(M \to Y\) relations. Second, the remaining \(M\to Y\) paths are estimated with regularization. Lastly, the test step performs the joint significance test as introduced by Baron \& Kenny (1986) with Bonferroni correction on the remaining mediators.

The main disadvantage of these methods is that there is a pertinent difference between (a) penalized estimation of the paths and (b) finding mediators. For XMed, a relatively small \(\alpha_p\) path will be shrunk to 0 before stronger \(\alpha\) paths, regardless of the strength of its associated \(\beta_p\) path. This holds for HIMA too, since in the selection stage it considers only \(\beta\) paths. Thus, these methods do not target \emph{mediators with strong indirect effects \(\alpha\beta\)}, but \emph{intermediate variables with strong \(\alpha\) or \(\beta\) paths}. Even though these methods do work conditionally, they make the implicit assumption that the mediators also have the strongest \(X\to M\) and \(M\to Y\) paths, which need not be so.

Rephrasing this in terms of decision functions, the regularization methods exclude variables which have a relatively weak covariance with \(X\) or \(Y\). However, this decision criterion only partially captures theoretically plausible mediators: true mediators may exist for which the covariance with \(X\) or \(Y\) is relatively weak, but the indirect effect \(\alpha\beta\) is relatively strong. The regularization methods will thus underperform in the presence of ``noise'' variables which are not mediators, but which strongly covary with either \(X\) or \(Y\). We illustrate this in the simulation section.

\vspace{.5cm}

In conclusion, to perform EMA, (a) the SEM method is optimal but unavailable for high-dimensional data, (b) the filter method is simple and flexible but does not select mediators conditionally, and (c) regularization methods do proper conditioning but are estimating paths rather than selecting mediators.

\hypertarget{coordinate-wise-mediation-filter}{%
\section{Coordinate-wise Mediation Filter}\label{coordinate-wise-mediation-filter}}

We propose a hybrid method, the Coordinate-wise Mediation Filter (CMF), which contains both theory-driven decision functions and conditional estimation of the quantity of interest. Like the filter method, CMF applies a decision function to each of the mediators, but it performs this task conditional on the set of currently selected mediators. The procedure is similar to cyclical coordinate descent, the algorithm underlying regularization procedures in various software implementations -- but differs in that mediation rather than separate regression paths are explicitly identified as the target. A key component of this algorithm is the use of residuals to remove dependency among the coordinates (Hastie et al., 2015). CMF generalizes this idea to mediator selection with arbitrary objective functions.

The CMF implementation consists of two components: an inner algorithm, which handles feature selection using the decision function \(\mathcal{D}\) through coordinate descent, and an outer algorithm, which performs random starts, feature subsampling, and subsequent aggregation. The combined procedure can be characterized as a stochastic coordinate descent algorithm. The following two sections give a detailed outline of the inner and outer algorithm.

\hypertarget{inner-algorithm}{%
\subsubsection{Inner algorithm}\label{inner-algorithm}}

First, we initialize a vector of length \(P\) which contains the current mediator selection in the form of 0 and 1 values -- the starting values. A \emph{step} is then as follows: for each potential mediator \(M_p\), create a data matrix \(\boldsymbol{M}_*\), which contains all the mediators currently selected, excluding the variable \(M_p\) under consideration. Then, perform the decision function \(\mathcal{D}\) on the parts of \(\boldsymbol{x}\) and \(\boldsymbol{y}\) orthogonal to (conditional on) this matrix. This conditioning is performed through calculating the \emph{residuals} of \(\boldsymbol{x}\) and \(\boldsymbol{y}\) with respect to \(\boldsymbol{M}_*\):

\[
\begin{aligned}
\boldsymbol{r}_{x} &= \boldsymbol{x} - \boldsymbol{M}_* (\boldsymbol{M}_*'\boldsymbol{M}_*)^{-1}\boldsymbol{M}_*'\boldsymbol{x}\nonumber\\
\boldsymbol{r}_{y} &= \boldsymbol{y} - \boldsymbol{M}_* (\boldsymbol{M}_*'\boldsymbol{M}_*)^{-1}\boldsymbol{M}_*'\boldsymbol{y}
\end{aligned} 
\]

The decision function is thus performed as \(\mathcal{D}(\boldsymbol{r}_{x}, \boldsymbol{M}_p, \boldsymbol{r}_{y})\), leading to a binary decision whether mediator \(p\) selected, conditional on \(\boldsymbol{M}_*\).

The inner algorithm is run continuously, randomly ordering the choice of \(p\) in each iteration. It stops either when the mediator selection does not change from one step to the next, or when the prespecified maximum number of iterations is reached. The resulting program, shown in Algorithm \ref{alg:cmf}, is a binary, randomized form of cyclical coordinate descent similar to those in Hastie et al. (2015). The randomization improves stability for very high-dimensional data (Nesterov, 2012). Richtárik \& Takáč (2014) show that this method attains relatively fast convergence even with a billion variables in a sparse regression situation.

\begin{algorithm}[H]
  \caption{Inner CMF algorithm}\label{alg:cmf}
  \begin{algorithmic}[1]
    \State \texttt{scale($\boldsymbol{x}$); scale($\boldsymbol{M}$); scale($\boldsymbol{y}$)} 
    \State \texttt{P $\gets$ ncol($\boldsymbol{M}$)}  \Comment{number of mediators}
    \State \texttt{decvec $\gets$  $0_1, 0_2, \ldots, 0_P$} \Comment{initialise 0/1 decision vector}
    \Repeat
      \For{\texttt{p in 1:P}}
      \State $\boldsymbol{M}_* \gets \boldsymbol{M}$\texttt{[, decvec \& !p]} \Comment{selected mediators excluding $p$}
      \State $\boldsymbol{r}_{x} \gets \boldsymbol{x} - \boldsymbol{M}_* (\boldsymbol{M}_*'\boldsymbol{M}_*)^{-1}\boldsymbol{M}_*'\boldsymbol{x}$ \Comment{residual of x}
      \State $\boldsymbol{r}_{y} \gets \boldsymbol{y} - \boldsymbol{M}_* (\boldsymbol{M}_*'\boldsymbol{M}_*)^{-1}\boldsymbol{M}_*'\boldsymbol{y}$ \Comment{residual of y}
      \State \texttt{decvec[p]} $\gets \mathcal{D}(\boldsymbol{r}_{x}, \boldsymbol{M}$\texttt{[, p]}$, \boldsymbol{r}_{y})$ \Comment{decision function}
      \EndFor
    \Until{\texttt{decvec == decvec}$_{prev}$  } \Comment{convergence when decvec is stable}
  \end{algorithmic}
\end{algorithm}

\hypertarget{outer-algorithm}{%
\subsubsection{Outer algorithm}\label{outer-algorithm}}

The value of the decision vector resulting from the inner algorithm depends to some extent on the starting values, due to the discrete nature of its coordinates. Therefore, the algorithm is embedded in an outer loop that performs multiple random starts. After aggregating the results from the different starts, the decision vector of length \(P\) is continuous: each element \(p\) in this vector signifies the proportion of times the potential mediator \(M_p\) was selected by the inner algorithm. These proportions, or \emph{empirical selection probabilities}, naturally lead to a mediator ranking. This ranking can then again be dichotomized using a cutoff score.

The second essential part in the outer algorithm is \emph{feature sampling}. With feature sampling, the inner algorithm will loop over only \(\big\lceil{\sqrt{P}}\big\rceil\) potential mediators at each iteration. This procedure is similar to how random forest decorrelates its trees (Breiman, 2001). Zhang, Zhao, Zhang, \& Wei (2017) show in a sparse regression setting that feature sampling improves and stabilizes the performance of feature selection. Furthermore, there are links between feature sampling and shrinkage: for linear regression, considering only \(\big\lceil{\sqrt{P}}\big\rceil\) variables during training is equivalent to ridge regression on the standardized predictors. This generalizes to more complex methods such as GLM (Wager, Wang, \& Liang, 2013). Feature sampling in the CMF algorithm thus takes on the crucial role of regularization.

The entire CMF procedure is implemented in the \texttt{R} package \texttt{cmfilter}, available from \href{https://github.com/vankesteren/cmfilter}{\texttt{https://github.com/vankesteren/cmfilter}}. An example analysis with specific hyperparameters and cutoff score determination is described in the application section to this paper, with accompanying \texttt{R} code in the supplementary material.

\vspace{.5cm}

The CMF method addresses the most important issues associated with both filter and regularization methods: it conditions on the other mediators while simultaneously being flexible to the choice of theoretically relevant decision functions. In the next section, we investigate the performance of CMF through simulation.

\hypertarget{simulations}{%
\section{Simulations}\label{simulations}}

This section is subdivided into two parts. The first part aims to show empirically the theoretical advantages and disadvantages of SEM, filter, XMed, HIMA, and CMF. We simulate specific conditions which are theoretically challenging for some but not all methods. The results from the first section are aimed at generating an understanding of the theoretical background in the present paper.

The second part is aimed at simulating real-world performance in a controlled high-dimensional situation. The results from this section indicate to what extent the CMF method outperforms its rival methods in practice, in addition to providing an anchor for the expected absolute level of performance in terms of false positives and true positives in such a situation.

All the simulations were run on R version 3.5.0 (R Core Team, 2018). The full environment used for the simulations is shown in Appendix \ref{app:envir}.

\hypertarget{theoretical-conditions}{%
\subsection{Theoretical conditions}\label{theoretical-conditions}}

The goal of this section is to illustrate when each method performs adequately and when it does not. Two situations are of particular interest: (a) suppression through correlation among mediators, and (b) noise in the \(\alpha\) and \(\beta\) paths, overshadowing a potential mediator. Filter methods are likely to underperform in terms of power in the first case, as the effect of a mediator is dependent on another and marginally invisible. In the second case, the regularization methods are theorized to under-perform because the \(\alpha\) and \(\beta\) paths are regularized independently whereas it is their combination that indicates mediation.

The data was controlled to behave according to the population, i.e., the data was transformed to exhibit the exact correlation matrix implied by the data-generating model. In each simulation, we show the power and false discovery rates of the three methods in 100 simulated datasets of 400-600 observations. The decision function under consideration for the filter, SEM, and CMF methods was the Sobel test (Sobel, 1986), one of the most common tests in the product of coefficients category (MacKinnon et al., 2002). For these tests, any variable with a \(p\)-value below .1 was considered to be a mediator. The SEM and filter methods were implemented using the \texttt{lavaan} package (Rosseeel, 2012), and CMF was implemented using the accompanying \texttt{cmfilter} package. For XMed, the \texttt{regsem} package (Jacobucci et al., 2016) was used with cross-validation was to find the optimal penalty parameter, and any variables with nonzero \(\alpha\) and \(\beta\) paths were considered mediators. HIMA was run according to its implementation in the R package \texttt{HIMA} (Zhang et al., 2016), again with a \(p\)-value of .1. Further details on the data generation and precise simulation conditions can be found in the the R code in the supplementary material.

\hypertarget{suppression}{%
\subsubsection{Suppression}\label{suppression}}

\begin{figure}[H]

{\centering \includegraphics[width=0.6\linewidth]{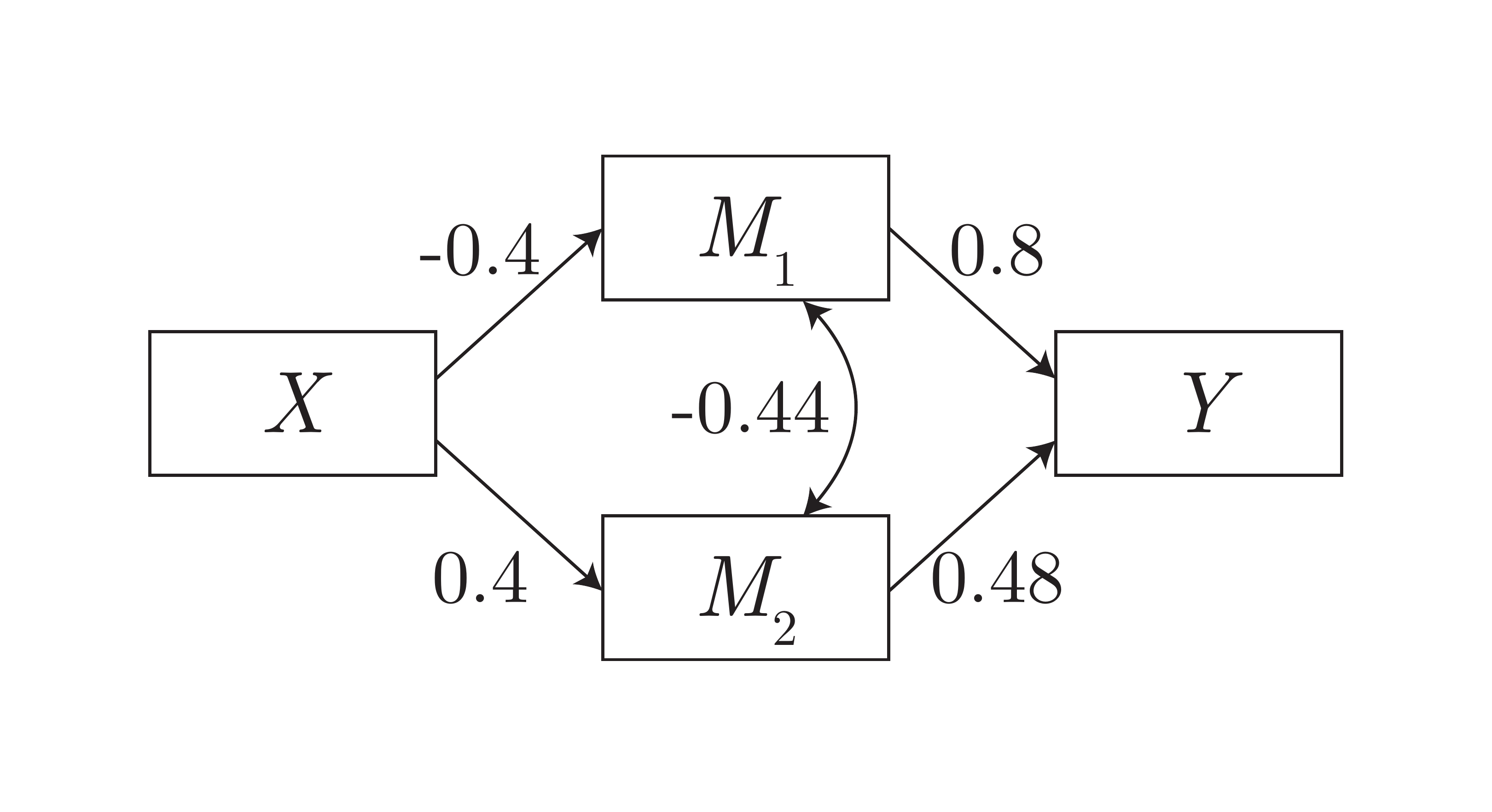} 

}

\caption{Data-generating model for the suppression simulation. Double-headed arrow indicates residual covariance.}\label{fig:supf}
\end{figure}

In the first illustration, the effect of the second \(\beta\) path is 0, but conditional on the first mediator this effect is nonzero. Its data-generating model is shown in Figure \ref{fig:supf}. The power to detect the second mediator thus indicates the robustness of each selection method to a full suppression effect.
\begin{align*}
&\text{cov}(M_1, M_2) = -0.44+-0.4\cdot0.4 = -0.6\\
&\text{cov}(M_2, Y) = 0.48+0.8\cdot \text{cov}(M_1, M_2) = 0
\end{align*}

The results are shown in Table \ref{tab:sup}, in the form of power to detect each mediator. As expected, the filter method fails to detect \(M_2\) under the marginal suppression in this data, whereas the other methods do detect the suppressed mediator.

\begin{table}[H]

\caption{\label{tab:sup}Empirical power, calculated as the proportions of selection for each mediator in the 100 generated datasets.}
\centering
\begin{tabular}{lcc}
\toprule
Method & $M_1$ & $M_2$\\
\midrule
SEM & 1 & 1\\
Filter & 1 & 0\\
XMed & 1 & 1\\
HIMA & 1 & 1\\
CMF & 1 & 1\\
\bottomrule
\end{tabular}
\end{table}

\hypertarget{noise-in-the-alpha-paths}{%
\subsubsection{\texorpdfstring{Noise in the \(\alpha\) paths}{Noise in the \textbackslash{}alpha paths}}\label{noise-in-the-alpha-paths}}

\begin{figure}[H]

{\centering \includegraphics[width=0.6\linewidth]{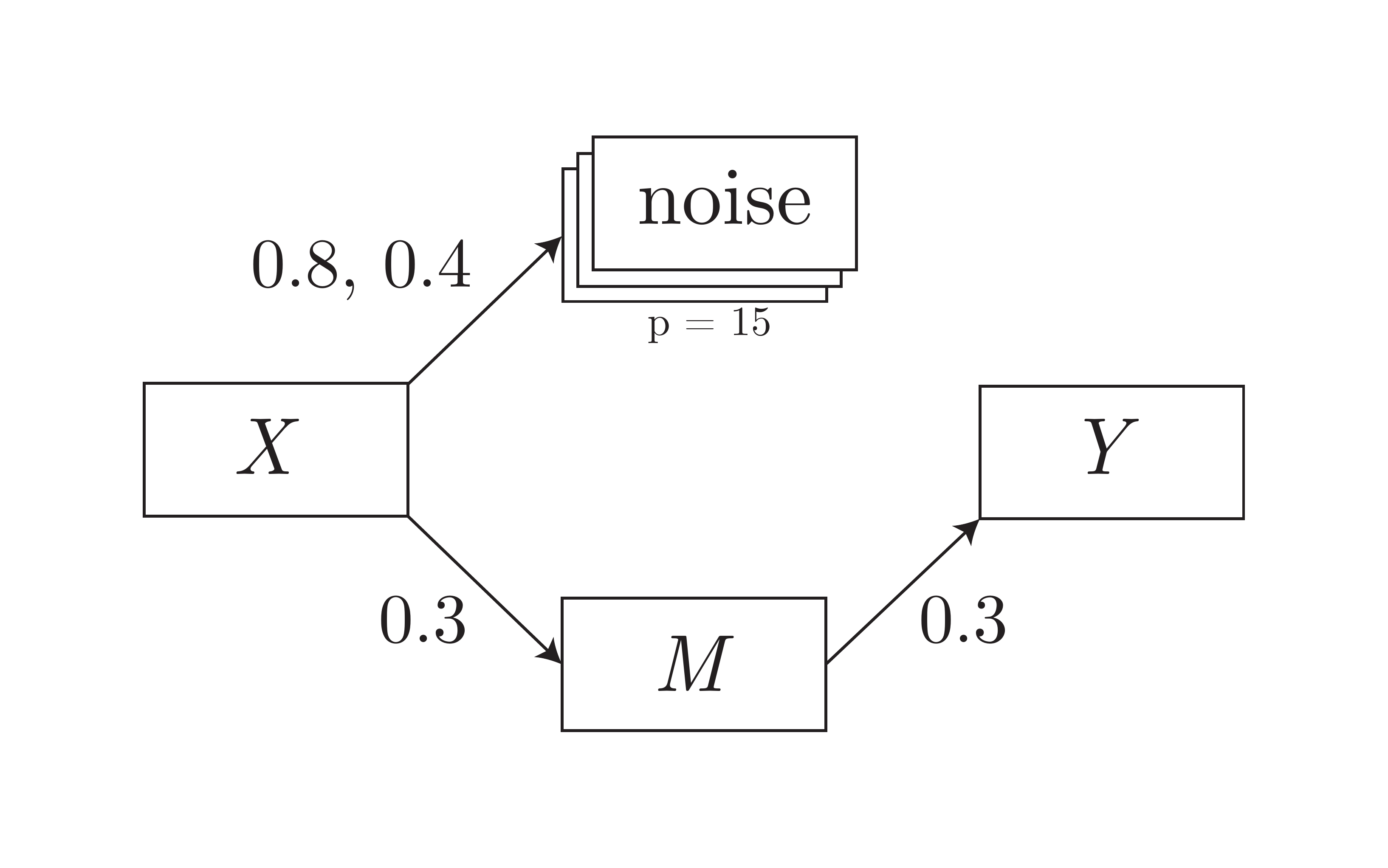} 

}

\caption{Data-generating model for the simulation of noise in the \(\alpha\) paths.}\label{fig:noisea}
\end{figure}

The second illustration considers noise in the form of variables related to \(X\). In addition to the single mediator, 15 noise variables were generated; the \(\alpha\) path was set to 0.8 for 3 of the variables, and 0.4 for the remaining 12. In addition, small residual correlations were induced in this set of variables to more closely resemble real-world patterns. The data-generating mechanism is shown in Figure \ref{fig:noisea}.

This situation challenges XMed, which considers the \(\alpha\) and \(\beta\) paths separately and is therefore theoretically more likely to select the strong paths rather than the mediating path, which has strength 0.3.

The results are displayed in Table \ref{tab:noia} in the form of rates of detection for each potential mediator. The SEM method performs optimally, as do the HIMA and CMF methods. The filter and XMed methods do not perform as well as these, having relatively strong false positive rates and lower power, respectively.

\begin{table}[H]

\caption{\label{tab:noia}Selection rates of each mediator in 100 simulated datasets where the noise variables (2-16) have a nonzero relation with the \(X\) variable. \(M\) is the true mediator, dot indicates 0.}
\centering
\begin{tabular}{lcccccccccccccccc}
\toprule
Method & $M$ & 2 & 3 & 4 & 5 & 6 & 7 & 8 & 9 & 10 & 11 & 12 & 13 & 14 & 15 & 16\\
\midrule
SEM & 99 & . & . & . & . & . & . & . & . & . & . & . & . & . & . & .\\
Filter & 100 & . & . & 100 & . & . & . & . & . & 100 & . & . & . & 76 & . & .\\
XMed & . & . & . & . & . & . & . & . & . & . & . & . & . & . & . & .\\
HIMA & 100 & . & . & . & . & . & . & . & . & . & . & . & . & . & . & .\\
CMF & 100 & . & . & . & . & . & . & . & . & . & . & . & . & . & . & .\\
\bottomrule
\end{tabular}
\end{table}

\hypertarget{noise-in-the-beta-paths}{%
\subsubsection{\texorpdfstring{Noise in the \(\beta\) paths}{Noise in the \textbackslash{}beta paths}}\label{noise-in-the-beta-paths}}

\begin{figure}[H]

{\centering \includegraphics[width=0.6\linewidth]{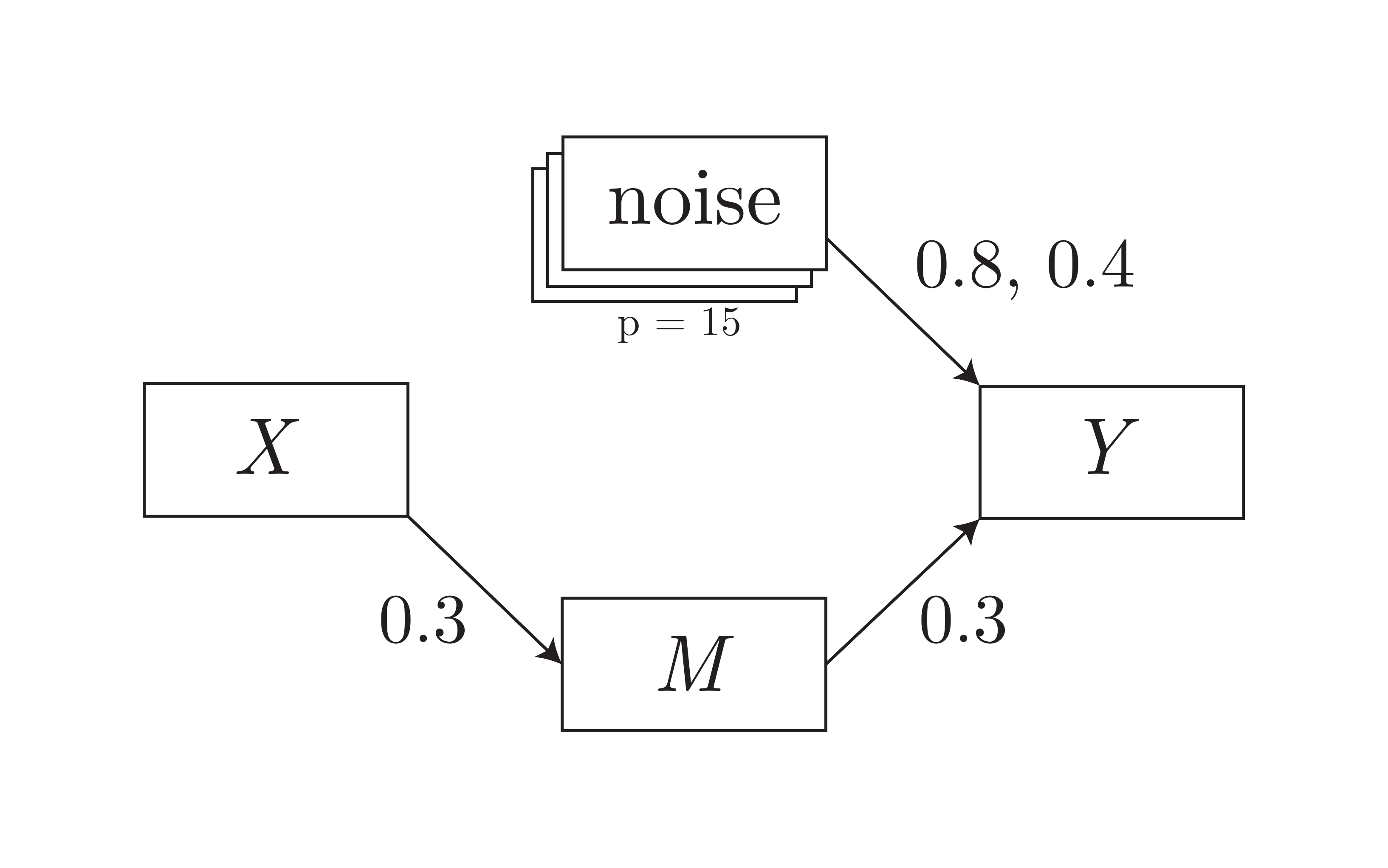} 

}

\caption{Data-generating model for the simulation of noise in the \(\beta\) paths.}\label{fig:noiseb}
\end{figure}

Like the second illustration, the third adds 15 noise variables alongside the true mediator. This time, the noise variables are related to the outcome variable \(Y\). The data-generating mechanism is shown in Figure \ref{fig:noiseb}.

The results can be found in Table \ref{tab:noib}. The HIMA method, which in the previous simulations performed as well as the benchmark SEM method, fails to detect the mediator in any of the 100 iterations. The other methods attain a perfect score.

\begin{table}[H]

\caption{\label{tab:noib}Selection rates of each mediator in 100 simulated datasets where the noise variables (2-16) have a nonzero relation with the \(Y\) variable. \(M\) is the true mediator, dot indicates 0.}
\centering
\begin{tabular}{lcccccccccccccccc}
\toprule
Method & $M$ & 2 & 3 & 4 & 5 & 6 & 7 & 8 & 9 & 10 & 11 & 12 & 13 & 14 & 15 & 16\\
\midrule
SEM & 99 & . & . & . & . & . & . & . & . & . & . & . & . & . & . & .\\
Filter & 100 & . & . & . & . & . & . & . & . & . & . & . & . & . & . & .\\
XMed & 92 & 5 & 5 & 6 & 6 & 3 & 4 & 2 & 3 & 4 & 2 & 3 & 4 & 5 & 6 & 3\\
HIMA & . & . & . & . & . & . & . & . & . & . & . & . & . & . & . & .\\
CMF & 100 & . & . & . & . & . & . & . & . & . & . & . & . & . & . & .\\
\bottomrule
\end{tabular}
\end{table}

\hypertarget{suppression-and-noise}{%
\subsubsection{Suppression and noise}\label{suppression-and-noise}}

\begin{figure}[H]

{\centering \includegraphics[width=0.6\linewidth]{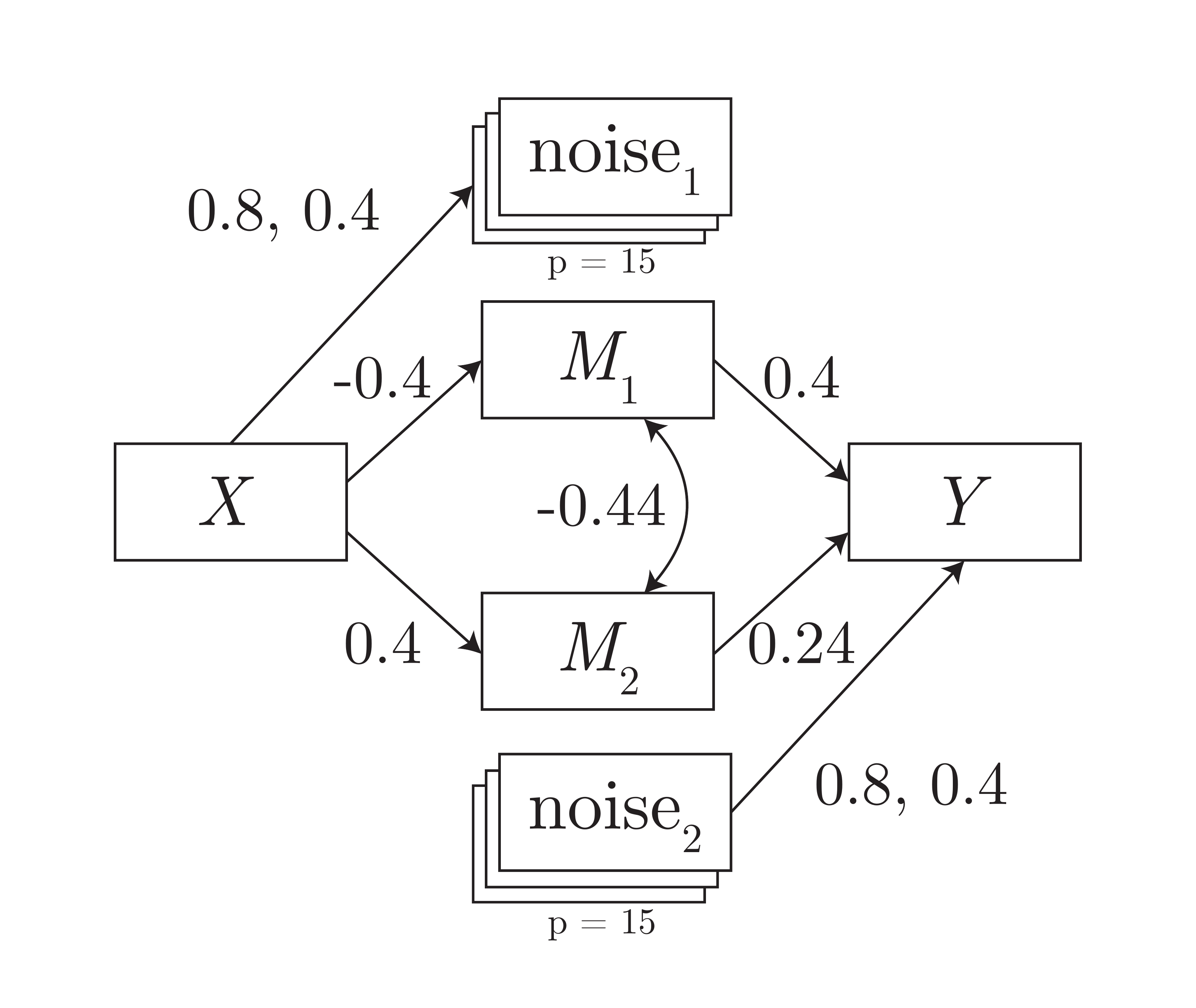} 

}

\caption{Data-generating model for the simulation of suppression with noise in the \(\alpha\) and \(\beta\) paths.}\label{fig:noisup}
\end{figure}

The last illustration combines the above simulations into a single data-generating mechanism, where both suppression and noise are present, as shown in Figure \ref{fig:noisup}.

The results of this combined simulation, displayed in Table \ref{tab:nois} show again that CMF performs at benchmark level. An interesting quantity for the imperfect methods is the positive predictive value (PPV): the probability that a mediator selected by a method truly mediates the effect of \(X\) on \(Y\). For filter and XMed methods, the PPV is lowered through either a relatively low true positive rate (power) or a high false positive rate (type-I error).

\begin{table}[H]

\caption{\label{tab:nois}True positive rates, false positive rates, and positive predictive values (PPV) of the combined suppression and noise simulation. The PPV indicates the probability that a mediator selected by the method is a true mediator.}
\centering
\begin{tabular}{lccrr}
\toprule
Method & Power $M_1$ & Power $M_2$ & FPR & PPV\\
\midrule
SEM & 0.99 & 0.99 & 0.000 & 1.00\\
Filter & 1.00 & 0.00 & 0.000 & 1.00\\
XMed & 0.88 & 0.87 & 0.097 & 0.37\\
HIMA & 1.00 & 0.00 & 0.000 & 1.00\\
CMF & 1.00 & 1.00 & 0.000 & 1.00\\
\bottomrule
\end{tabular}
\end{table}

\hypertarget{interim-conclusion}{%
\subsubsection{Interim conclusion}\label{interim-conclusion}}

While the considered data-generating mechanisms are very specific, the differences in performance between the methods can be exacerbated and diminished by altering the parameter values while preserving the structure. Overall, CMF is the only method that performs as well as the baseline in all of these data-generating mechanisms. Together, they show that this method is robust to boundary cases where other methods may fail. This is a valuable property of a mediator selection method, because these situations may occur simultaneously, with no way to test them in real-world datasets. In the next part, we explore how well the CMF method performs in high-dimensional circumstances, where the baseline optimal SEM method cannot work.

\hypertarget{high-dimensional-mediation-simulation}{%
\subsection{High-dimensional mediation simulation}\label{high-dimensional-mediation-simulation}}

In this section, we compare the performance of the available EMA methods in a simplified high-dimensional situation. Due to the wide nature of the dataset (\(p = 1000\)), the benchmark default SEM method is unavailable.

\hypertarget{simulation-setup}{%
\subsubsection{Simulation setup}\label{simulation-setup}}

\begin{figure}[H]

{\centering \includegraphics[width=0.6\linewidth]{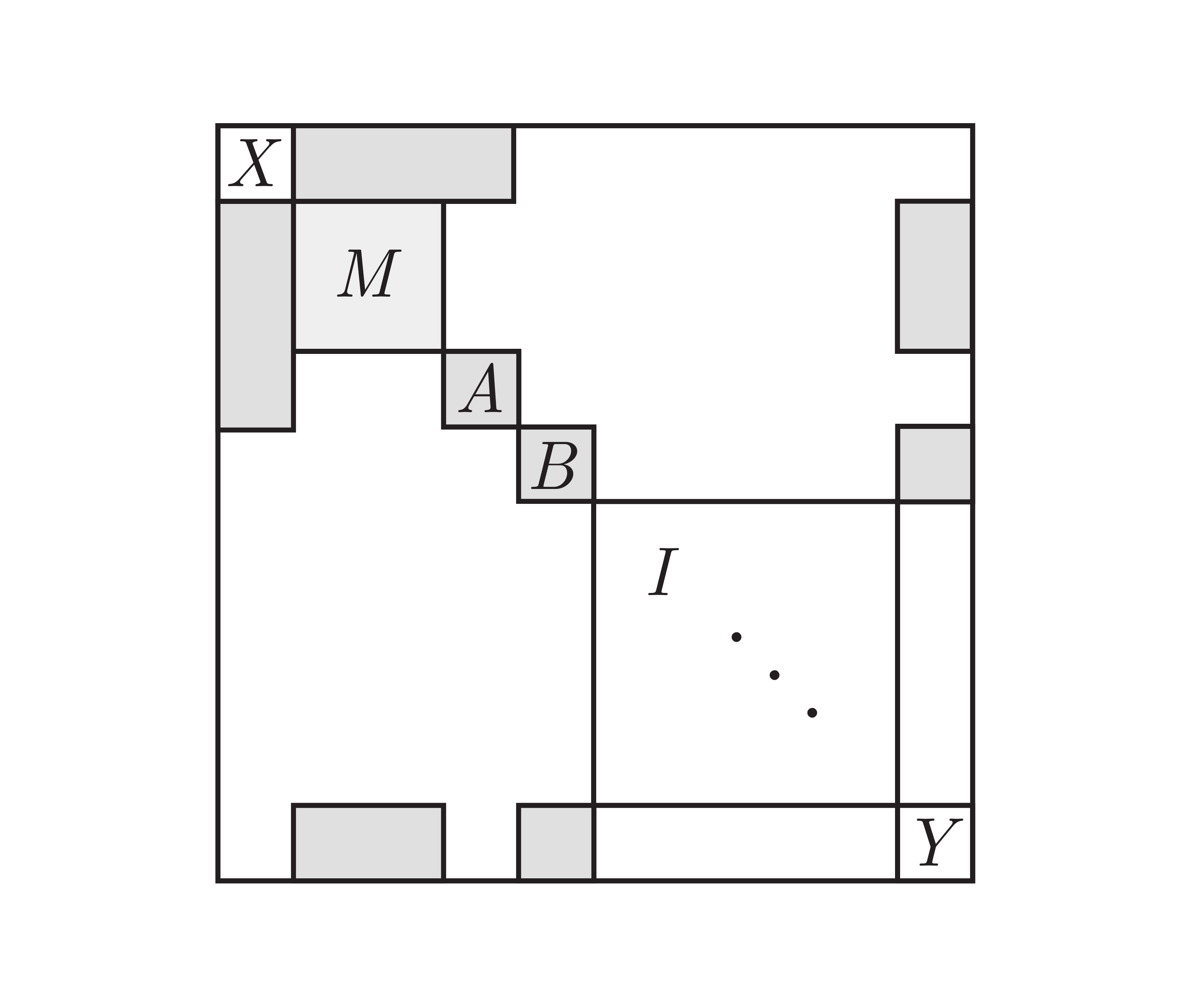} 

}

\caption{General covariance structure for the high-dimensional performance simulation. In the white sections of the matrix, there is no covariance. The true mediator block \(M\) is related to both \(X\) and \(Y\), whereas the correlating noise blocks are related to either \(X\) (block \(A\)) or \(Y\) (block \(B\)). The largest block is the identity matrix block \(I\), which generates only unrelated noise variables.}\label{fig:mondriaan}
\end{figure}

Following one of the high-dimensional simulation conditions of Zhang et al. (2016), the dataset consists of 100 samples and 1000 potential mediators. These mediators are generated in four uncorrelated blocks: one block with true mediators (\(M\)), one with noise variables related to \(X\) (\(A\)), one noise block covarying with \(Y\) (\(B\)), and one large ``white noise'' block without any covariance (\(I\)). The general structure can be found in Figure \ref{fig:mondriaan}. For each of the simulations, this structure was created as a sparse block matrix using the \texttt{Matrix} package (Bates \& Maechler, 2017), after which multivariate normal data was generated using the \texttt{sparseMVN} package (Braun, 2018). Specific data generation and simulation R code can be found in the supplementary materials.

Note that unlike the illustrative simulations, these data favor the filter method: there is no suppression or excessive interdependence of potential mediators. Therefore, the filter method is the benchmark in this simulation. The XMed method was omitted from this simulation because it requires estimation of the full SEM model before regularizing: it would need to be adjusted to work with high-dimensional data.

\hypertarget{results}{%
\subsubsection{Results}\label{results}}

The results are displayed in Table \ref{tab:hidim}. The CMF method has the highest true positive rate, and a medium false positive rate, leading to a similar positive predictive value (PPV) to the filter method. In other words, the mediators selected by CMF are as likely to be true mediators as those selected by the benchmark filter method. As true positive rates and false positive rates can be adjusted by the choice of alpha level, we conclude that the CMF method also performs at benchmark level in this high-dimensional situation.

\begin{table}[H]

\caption{\label{tab:hidim}True positive rates, false positive rates, and positive predictive values for the high-dimensional data simulation. Note that XMed failed to run as-is for the simulated datasets, as it required running the full SEM model before regularizing.}
\centering
\begin{tabular}{llcc}
\toprule
  & Power & Type I Error & PPV\\
\midrule
CMF & 0.2648 & 0.00258 & 0.5068\\
Filter & 0.2412 & 0.00235 & 0.5124\\
HIMA & 0.0686 & 0.00941 & 0.0323\\
\bottomrule
\end{tabular}
\end{table}

\hypertarget{application-to-epigenetic-data}{%
\section{Application to epigenetic data}\label{application-to-epigenetic-data}}

In this section, we show how the CMF method can be used for exploratory mediation analysis in a real-world setting. Aside from the results shown here, the full \texttt{R} syntax is available in the supplementary material.

Houtepen et al. (2016) researched which locations in the genome are likely to mediate the relation between childhood trauma and stress reactivity later in life. In order to identify the genomic locations, they measured methylation at CpG sites using array based technology. In a discovery sample, they found a location of interest which they subsequently researched further and related to functional changes in the human prefrontal cortex.

Here, we re-analyze the original discovery sample dataset to investigate whether CMF yields different potentially relevant locations compared to the correlational filter analysis of the original authors.

\hypertarget{dataset-and-preprocessing}{%
\subsubsection{Dataset and preprocessing}\label{dataset-and-preprocessing}}

The dataset of the discovery sample was obtained from ArrayExpress, the data repository of the European Bioinformatics Institute: \url{https://www.ebi.ac.uk/arrayexpress/experiments/E-GEOD-77445}. The sample consists of 85 healthy individuals. The \(X\) variable is score on a childhood trauma questionnaire and the \(Y\) variable is the increase in cortisol after a stress test defined as increase in the area under the curve (iAUC). The 385 884 potential mediators \(M\) were taken from the analysis of DNA methylation in the blood, with default preprocessing. From the available respondent characteristics, age and sex were considered to be confounders. For full details of the dataset, see Houtepen et al. (2016).

Before analysis, \(X\), \(Y\), and \(M\) were residualized with respect to their intercept, age, and sex. Since the number of \(M\) variables was so large, the last preprocessing step was a straightforward univariate filter. For this, the top 1000 potential mediators in terms of their absolute product of correlations with \(X\) and \(Y\) were retained. For more details, see the preprocessing R code in the supplementary materials.

\hypertarget{analysis-and-results}{%
\subsubsection{Analysis and Results}\label{analysis-and-results}}

\begin{figure}[H]

{\centering \includegraphics[width=1\linewidth]{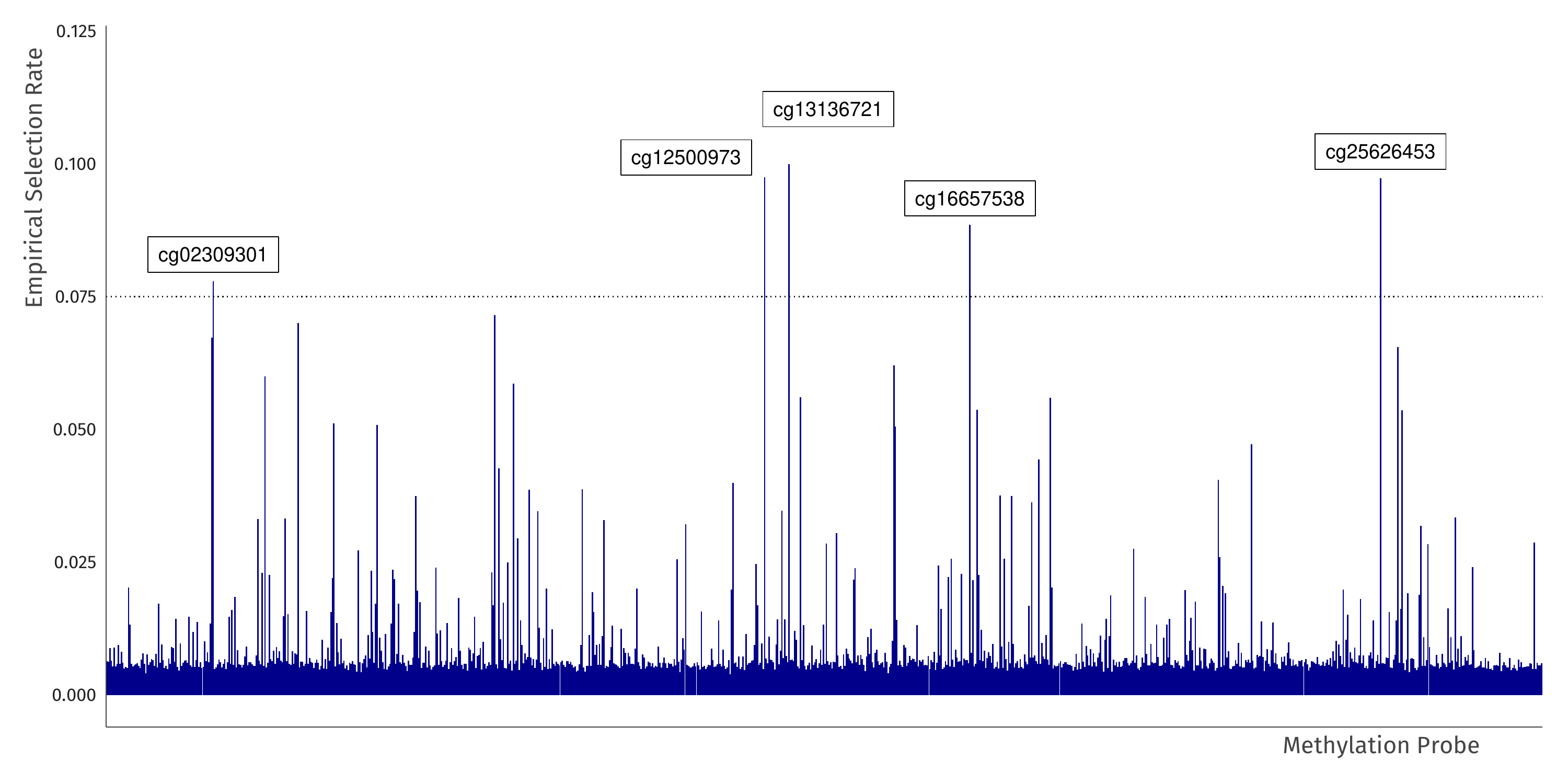} 

}

\caption{Selection rates of the potential mediators in the methylation dataset.}\label{fig:selrate}
\end{figure}

The CMF algorithm was performed using the centered \(X\) and \(Y\) and the 1000 potential mediators \(M\). The Sobel test with a \(p\)-value of 0.1 was used as the decision function \(\mathcal{D}\) and 10 000 iterations with random starts were run to ensure stability of the results. After inspecting the scree plot of the selection rates, the cutoff for selection was set to 0.075. The resulting selection rates and selected \texttt{cg} locations in the genome are shown in Figure \ref{fig:selrate}.

These locations were annotated using the BioConductor package \texttt{FDb.InfiniumMethylation.hg18} (Triche, 2014) to find the nearest protein-coding gene. The shortened descriptions were summarized from the GeneCards database (Safran et al., 2002). The result is shown in Table \ref{tab:gene}.

\begin{table}[H]

\caption{\label{tab:gene}Annotation of the selected mediators from the CMF algorithm.}
\centering
\begin{tabular}{lll}
\toprule
Probe & Gene & Description\\
\midrule
cg16657538 & ZSCAN30 & Involved in transcriptional regulation\\
cg25626453 & PRRC2A & Associated with the age-at-onset of diabetes\\
cg02309301 & ARGLU1 & Associated with sexual development\\
cg13136721 & RPTOR & Involved in regulation of cell growth and survival\\
cg12500973 & HNRNPF & Involved in regulation of mRNA\\
\bottomrule
\end{tabular}
\end{table}

Inspecting and comparing these results more closely, two of the locations identified by CMF have been previously associated with development throughout the lifespan: PRRC2A and ARGLU1. These two locations are also in the top 10 of the lists generated by HIMA and filter. In addition, the RPTOR gene has been associated with cell growth and survival -- development on a cellular level. Relative to other sites, this last location does not have a strong correlation with either childhood trauma (\emph{r} = 0.186) or stress reactivity (\emph{r} = 0.233), but due to its conditional indirect effect it is deemed relevant by both CMF and HIMA. The ZSCAN30 gene has a small marginal correlation with stress reactivity (\emph{r} = 0.096) which lowers its rank for both the filter and HIMA methods. However, due to its strong correlation with childhood trauma (\emph{r} = 0.347) and its conditional relevance this site is still high on the list for CMA.

In conclusion, CMA has overlap with other methods but can identify relevant locations that other methods may miss. Further research using replication samples could focus on exploring whether and how methylation at these locations may alter stress reactivity after childhood trauma.

\hypertarget{discussion}{%
\section{Discussion}\label{discussion}}

Structural equation modeling, the benchmark method for exploratory mediation analysis, is unavailable in the case of high-dimensional data. Several alternative methods exist, but in the current paper we have shown through simulations that these underperform in situations with specific dependence among mediators, noise variables related to either \(X\) or \(Y\), or a combination thereof. Taking these situations into account, we have introduced CMF, a hybrid algorithmic method to identify from a set of potential mediators the most likely true mediators.

CMF improves upon the existing methods by combining the estimation method from regularized regression with the theory-based decision functions from classic mediation analysis. It extends EMA with theoretically relevant decision functions to the high-dimensional case. As a full package including a software implementation, it is flexible to the choice of decision function, robust in the tested situations, and it scales to multiple processor cores.

Besides its role as a novel method for EMA, CMF contributes several ideas to the statistical literature. It shows that the use of cyclically calculated residuals is applicable beyond regression into the territory of structural equation modeling. In addition, its performance is greatly improved by feature subsampling, which has regularizing effects on the estimated parameters and thus on the mediator selections. CMF is an example of how combining a deterministic algorithm with a stochastic outer component can lead to adequate performance.

One result of the approach taken in this paper is that there is no formal proof of convergence, and the algorithm may take a long time to stabilize. In addition, the complications introduced in the outer loop make determination of the cutoff for selection nontrivial. In general, the algorithm will output a top-N vector of most selected mediators, and potential options for deciding which cutoff to take are visual inspection of the scree plot or a form of parallel analysis (Horn, 1965). In addition, the error rates (type I and type II errors) are not analytically defined and have a complex relation with the alpha level of the base decision function. This could be investigated empirically in the future.

For this work, we only considered direct feature selection on the set of \(M\) variables. Another solution is projecting the available features onto a low-dimensional space before or during estimation. Feature selection can then be performed in this space, leading to variable importance upon reprojection to the original space. Examples are PCA, PLS, or the directions of mediation method by (Chén, Crainiceanu, Ogburn, Caffo, \& Wager, 2017). However, we chose to exclude these methods because they do not select mediators, but rather linear combinations of all mediators.

Our coordinate-wise mediation filter bears resemblance to a class of metaheuristic algorithms in the SEM literature for specification search (Marcoulides \& Falk, 2018). These algorithms perform an exploratory search for the optimal model based on overall model fit, e.g., the BIC objective. CMF could be considered specification search where the objective is not overall model fit but mediation analysis: it is targeted towards determining whether a specific variable is relevant to a process rather than searching for the optimal model. In addition, CMF performs regularization required for high-dimensional data. In the future, other specification search strategies could be implemented for EMA, but they each need to be adjusted to incorporate both a specific mediation objective and regularization.

Future research should focus on embedding mediation analysis theory directly in penalization procedures for these datasets, either in a classical estimation setting (Zhao \& Luo, 2016) or using Bayesian estimation with shrinkage priors (Erp, Oberski, \& Mulder, 2018). More generally, enriching structural equation models beyond EMA with embedded feature selection mechanisms will enable social and behavioral scientists to develop and test theories on novel, high-dimensional datasets.

\hypertarget{funding-details}{%
\section{Funding details}\label{funding-details}}

This work was supported by the Netherlands Organization for Scientific Research (NWO) under Grant number 406.17.057.

\hypertarget{acknowledgments}{%
\section{Acknowledgments}\label{acknowledgments}}

We thank Marco Boks, Yves Rosseel, Katrijn van Deun, Milica \(\text{Mio\v{c}evi\'{c}}\), and Ayoub Bagheri for their helpful suggestions at various stages of this work.

\newpage

\hypertarget{references}{%
\section{References}\label{references}}

\begingroup
\setlength{\parindent}{-0.5in}
\setlength{\leftskip}{0.5in}

\noindent

\hypertarget{refs}{}
\leavevmode\hypertarget{ref-Ammerman2018}{}%
Ammerman, B. A., Serang, S., Jacobucci, R., Burke, T. A., Alloy, L. B., \& McCloskey, M. S. (2018). Exploratory analysis of mediators of the relationship between childhood maltreatment and suicidal behavior. \emph{Journal of Adolescence}, \emph{69}, 103--112.

\leavevmode\hypertarget{ref-Atlas2014}{}%
Atlas, L. Y., Lindquist, M. A., Bolger, N., \& Wager, T. D. (2014). Brain mediators of the effects of noxious heat on pain. \emph{Pain}, \emph{155}(8), 1632--1648. \url{https://doi.org/10.1016/j.pain.2014.05.015}

\leavevmode\hypertarget{ref-Baron1986a}{}%
Baron, R. M., \& Kenny, D. a. (1986). The Moderator-Mediator Variable Distinction in Social The Moderator-Mediator Variable Distinction in Social Psychological Research: Conceptual, Strategic, and Statistical Considerations. \emph{Journal of Personality and Social Psychology}, \emph{51}(6), 1173--1182. \url{https://doi.org/10.1037/0022-3514.51.6.1173}

\leavevmode\hypertarget{ref-Bates2017}{}%
Bates, D., \& Maechler, M. (2017). Matrix: Sparse and Dense Matrix Classes and Methods. Retrieved from \url{https://cran.r-project.org/package=Matrix}

\leavevmode\hypertarget{ref-Boca2014}{}%
Boca, S. M., Sinha, R., Cross, A. J., Moore, S. C., \& Sampson, J. N. (2014). Testing multiple biological mediators simultaneously. \emph{Bioinformatics}, \emph{30}(2), 214--220. \url{https://doi.org/10.1093/bioinformatics/btt633}

\leavevmode\hypertarget{ref-Braun2018}{}%
Braun, M. (2018). sparseMVN: Multivariate Normal Functions for Sparse Covariance and Precision Matrices. R package version. Retrieved from \url{https://cran.r-project.org/package=sparseMVN}

\leavevmode\hypertarget{ref-Breheny2011}{}%
Breheny, P., \& Huang, J. (2011). Coordinate descent algorithms for nonconvex penalized regression, with applications to biological feature selection. \emph{Annals of Applied Statistics}, \emph{5}(1), 232--253. \url{https://doi.org/10.1214/10-AOAS388}

\leavevmode\hypertarget{ref-Breiman2001}{}%
Breiman, L. (2001). Random forests. \emph{Machine Learning}, \emph{45}(1), 5--32. \url{https://doi.org/10.1023/A:1010933404324}

\leavevmode\hypertarget{ref-Chen2017}{}%
Chén, O. Y., Crainiceanu, C., Ogburn, E. L., Caffo, B. S., \& Wager, T. O. R. D. (2017). High-dimensional multivariate mediation with application to neuroimaging data. \emph{Biostatistics}, (September), 1--16. \url{https://doi.org/10.1093/biostatistics/kxx040}

\leavevmode\hypertarget{ref-vanErp2018}{}%
Erp, S. van, Oberski, D. L., \& Mulder, J. (2018). Shrinkage Priors for Bayesian Penalized Regression. \emph{OSF Preprint}, 1--39. \url{https://doi.org/10.31219/osf.io/cg8fq}

\leavevmode\hypertarget{ref-Friedman2009}{}%
Friedman, J., Hastie, T., \& Tibshirani, R. (2009). Regularization Paths for Generalized Linear Models via Coordinate Descent. \emph{Journal of Statistical Software}, \emph{33}(1), 1--24.

\leavevmode\hypertarget{ref-Guyon2003}{}%
Guyon, I., \& Elisseeff, A. (2003). An Introduction to Variable and Feature Selection. \emph{Journal of Machine Learning Research (JMLR)}, \emph{3}(3), 1157--1182. \url{https://doi.org/10.1016/j.aca.2011.07.027}

\leavevmode\hypertarget{ref-Hastie2015}{}%
Hastie, T., Tibshirani, R., \& Wainwright, M. (2015). \emph{Statistical Learning with Sparsity: The Lasso and Generalizations} (p. 362). Boca Raton: CRC Press. \url{https://doi.org/10.1201/b18401-1}

\leavevmode\hypertarget{ref-Hayes2010}{}%
Hayes, A. F., \& Preacher, K. J. (2010). Quantifying and Testing Indirect Effects in Simple Mediation Models When the Constituent Paths Are Nonlinear. \emph{Multivariate Behavioral Research}, \emph{45}(4), 627--660. \url{https://doi.org/10.1080/00273171.2010.498290}

\leavevmode\hypertarget{ref-Hayes2014}{}%
Hayes, A. F., \& Preacher, K. J. (2014). Statistical mediation analysis with a multicategorical independent variable. \emph{British Journal of Mathematical and Statistical Psychology}, \emph{67}, 451--470. \url{https://doi.org/10.1111/bmsp.12028}

\leavevmode\hypertarget{ref-Horn1965}{}%
Horn, J. (1965). A Rationale and Test for the Number of Factors in Factor Analysis. \emph{Psychometrika}, \emph{30}(2), 179--185.

\leavevmode\hypertarget{ref-Houtepen2016}{}%
Houtepen, L. C., Vinkers, C. H., Carrillo-Roa, T., Hiemstra, M., Lier, P. A. van, Meeus, W., \ldots{} Boks, M. P. M. (2016). Genome-wide DNA methylation levels and altered cortisol stress reactivity following childhood trauma in humans. \emph{Nature Communications}, \emph{7}, 10967. \url{https://doi.org/10.1038/ncomms10967}

\leavevmode\hypertarget{ref-Jacobucci2018a}{}%
Jacobucci, R., Brandmaier, A. M., \& Kievit, R. A. (2018). Variable Selection in Structural Equation Models with Regularized MIMIC Models. \emph{PsyArXiv Preprint}, 1--40. \url{https://doi.org/10.17605/OSF.IO/BXZJF}

\leavevmode\hypertarget{ref-Jacobucci2016}{}%
Jacobucci, R., Grimm, K. J., \& McArdle, J. J. (2016). Regularized Structural Equation Modeling. \emph{Structural Equation Modeling}, \emph{23}(4), 555--566. \url{https://doi.org/10.1080/10705511.2016.1154793.Regularized}

\leavevmode\hypertarget{ref-Liu2013}{}%
Liu, Y., Aryee, M. J., Padyukov, L., Fallin, M. D., Hesselberg, E., Runarsson, A., \ldots{} Feinberg, A. P. (2013). Epigenome-wide association data implicate DNA methylation as an intermediary of genetic risk in rheumatoid arthritis. \emph{Nature Biotechnology}, \emph{31}(2), 142--147. \url{https://doi.org/10.1038/nbt.2487}

\leavevmode\hypertarget{ref-Mackinnon2008}{}%
MacKinnon, D. P. (2008). \emph{Introduction to statistical mediation analysis}. Routledge.

\leavevmode\hypertarget{ref-MacKinnon2007}{}%
MacKinnon, D. P., Fairchild, A. J., \& Fritz, M. S. (2007). Mediation Analysis. \emph{Annual Review of Psychology}, \emph{58}(1), 593--614. \url{https://doi.org/10.1146/annurev.psych.58.110405.085542}

\leavevmode\hypertarget{ref-MacKinnon2002}{}%
MacKinnon, D. P., Lockwood, C. M., Hoffman, J. M., West, S. G., \& Sheets, V. (2002). A comparison of methods to test mediation and other intervening variable effects. \emph{Psychological Methods}, \emph{7}(1), 83--104.

\leavevmode\hypertarget{ref-MacKinnon2004}{}%
MacKinnon, D. P., Lockwood, C. M., \& Williams, J. (2004). Confidence Limits for the Indirect Effect: Distribution of the Product and Resampling Methods. \emph{Multivariate Behavioral Research}, \emph{39}(1), 99--128. \url{https://doi.org/10.1207/s15327906mbr3901}

\leavevmode\hypertarget{ref-Marcoulides2018}{}%
Marcoulides, K. M., \& Falk, C. F. (2018). Model specification searches in structural equation modeling with r. \emph{Structural Equation Modeling: A Multidisciplinary Journal}, \emph{25}(3), 484--491.

\leavevmode\hypertarget{ref-Nesterov2012}{}%
Nesterov, Y. (2012). Efficiency of Coordinate Descent Methods on Huge-Scale Optimization Problems. \emph{SIAM J. Optim.}, \emph{22}(2), 341--362.

\leavevmode\hypertarget{ref-Olkin1995}{}%
Olkin, I., \& Finn, J. D. (1995). Correlations redux. \emph{Psychological Bulletin}, \emph{118}(1), 155--164. \url{https://doi.org/10.1037/0033-2909.118.1.155}

\leavevmode\hypertarget{ref-Preacher2015}{}%
Preacher, K. J. (2015). Advances in Mediation Analysis: A Survey and Synthesis of New Developments. \emph{Annual Review of Psychology}, \emph{66}(1), 825--852. \url{https://doi.org/10.1146/annurev-psych-010814-015258}

\leavevmode\hypertarget{ref-Preacher2008}{}%
Preacher, K. J., \& Hayes, A. F. (2008). Asymptotic and resampling strategies for assessing and comparing indirect effects in multiple mediator models. \emph{Behavior Research Methods}, \emph{40}(3), 879--891. \url{https://doi.org/10.3758/BRM.40.3.879}

\leavevmode\hypertarget{ref-RCoreTeam2018}{}%
R Core Team. (2018). R: A language and environment for statistical computing. Vienna, Austria: R Foundation for Statistical Computing. Retrieved from \url{https://www.r-project.org/}

\leavevmode\hypertarget{ref-Richtarik2014}{}%
Richtárik, P., \& Takáč, M. (2014). Iteration complexity of randomized block-coordinate descent methods for minimizing a composite function. \emph{Mathematical Programming}, \emph{144}(1), 1--38. \url{https://doi.org/10.1007/s10107-012-0614-z}

\leavevmode\hypertarget{ref-Rosseel2012}{}%
Rosseeel, Y. (2012). Lavaan: An R package for structural equation modeling. \emph{Journal of Statistical Software}, \emph{48}(2), 1--36.

\leavevmode\hypertarget{ref-Safran2002}{}%
Safran, M., Solomon, I., Shmueli, O., Lapidot, M., Shen-Orr, S., Adato, A., \ldots{} Lancet, D. (2002). GeneCards™ 2002: Towards a complete, object-oriented, human gene compendium. \emph{Bioinformatics}, \emph{18}, 1542--1543. \url{https://doi.org/10.1093/bioinformatics/18.11.1542}

\leavevmode\hypertarget{ref-Serang2017}{}%
Serang, S., Jacobucci, R., Brimhall, K. C., \& Grimm, K. J. (2017). Exploratory Mediation Analysis via Regularization. \emph{Structural Equation Modeling}, \emph{24}(5), 733--744. \url{https://doi.org/10.1080/10705511.2017.1311775}

\leavevmode\hypertarget{ref-Sobel1986}{}%
Sobel, M. E. (1986). Some new results on indirect effects and their standard errors in covariance structure models. \emph{Sociological Methodology}, \emph{16}(16), 159--186.

\leavevmode\hypertarget{ref-Triche2014}{}%
Triche, T. (2014). FDb.InfiniumMethylation.hg18: Annotation package for Illumina Infinium DNA methylation probes.

\leavevmode\hypertarget{ref-VanderWeele2015}{}%
VanderWeele, T. J. (2015). \emph{Explanation in Causal Inference: Methods for Mediation and Interaction} (p. 706). New York: Oxford University Press.

\leavevmode\hypertarget{ref-Vanderweele2014}{}%
Vanderweele, T. J., \& Vansteelandt, S. (2014). Mediation Analysis with Multiple Mediators. \emph{Epidemiologic Methods}, \emph{2}(1), 95--115. \url{https://doi.org/10.1515/em-2012-0010.Mediation}

\leavevmode\hypertarget{ref-wager2013}{}%
Wager, S., Wang, S., \& Liang, P. S. (2013). Dropout training as adaptive regularization. In \emph{Advances in neural information processing systems} (pp. 351--359).

\leavevmode\hypertarget{ref-Zhang2016}{}%
Zhang, H., Zheng, Y., Zhang, Z., Gao, T., Joyce, B., Yoon, G., \ldots{} Liu, L. (2016). Estimating and testing high-dimensional mediation effects in epigenetic studies. \emph{Bioinformatics}, \emph{32}(20), 3150--3154. \url{https://doi.org/10.1093/bioinformatics/btw351}

\leavevmode\hypertarget{ref-Zhang2017}{}%
Zhang, J., Zhao, Z., Zhang, K., \& Wei, Z. (2017). A feature sampling strategy for analysis of high dimensional genomic data. \emph{IEEE/ACM Transactions on Computational Biology and Bioinformatics}.

\leavevmode\hypertarget{ref-Zhao2016}{}%
Zhao, Y., \& Luo, X. (2016). \emph{Pathway Lasso: Estimate and Select Sparse Mediation Pathways with High Dimensional Mediators}. Retrieved from \url{http://arxiv.org/abs/1603.07749}

\endgroup

\newpage

\hypertarget{appendix-appendix}{%
\appendix}

\hypertarget{app:envir}{%
\section{R environment used}\label{app:envir}}

\begin{verbatim}
## R version 3.5.0 (2018-04-23)
## Platform: x86_64-w64-mingw32/x64 (64-bit)
## Running under: Windows >= 8 x64 (build 9200)
## 
## Matrix products: default
## 
## locale:
## [1] LC_COLLATE=Dutch_Netherlands.1252  LC_CTYPE=Dutch_Netherlands.1252   
## [3] LC_MONETARY=Dutch_Netherlands.1252 LC_NUMERIC=C                      
## [5] LC_TIME=Dutch_Netherlands.1252    
## 
## attached base packages:
## [1] stats4    parallel  stats     graphics  grDevices utils     datasets 
## [8] methods   base     
## 
## other attached packages:
##  [1] Massign_1.1.0                          
##  [2] firatheme_0.1.0                        
##  [3] FDb.InfiniumMethylation.hg18_2.2.0     
##  [4] org.Hs.eg.db_3.6.0                     
##  [5] TxDb.Hsapiens.UCSC.hg18.knownGene_3.2.2
##  [6] pbapply_1.3-4                          
##  [7] MASS_7.3-50                            
##  [8] HIMA_1.0.7                             
##  [9] ncvreg_3.10-0                          
## [10] regsem_1.1.2                           
## [11] Rcpp_0.12.18                           
## [12] lavaan_0.6-2                           
## [13] cmfilter_0.2.1                         
## [14] magrittr_1.5                           
## [15] forcats_0.3.0                          
## [16] stringr_1.3.1                          
## [17] dplyr_0.7.6                            
## [18] purrr_0.2.5                            
## [19] readr_1.1.1                            
## [20] tidyr_0.8.1                            
## [21] tibble_1.4.2                           
## [22] ggplot2_3.0.0                          
## [23] tidyverse_1.2.1                        
## [24] GenomicFeatures_1.32.1                 
## [25] AnnotationDbi_1.42.1                   
## [26] Biobase_2.40.0                         
## [27] GenomicRanges_1.32.6                   
## [28] GenomeInfoDb_1.16.0                    
## [29] IRanges_2.14.10                        
## [30] S4Vectors_0.18.3                       
## [31] BiocGenerics_0.26.0                    
## [32] glmnet_2.0-16                          
## [33] foreach_1.4.4                          
## [34] Matrix_1.2-14                          
## 
## loaded via a namespace (and not attached):
##  [1] nlme_3.1-137                bitops_1.0-6               
##  [3] matrixStats_0.54.0          lubridate_1.7.4            
##  [5] bit64_0.9-7                 doParallel_1.0.11          
##  [7] progress_1.2.0              httr_1.3.1                 
##  [9] rprojroot_1.3-2             tools_3.5.0                
## [11] backports_1.1.2             R6_2.2.2                   
## [13] DBI_1.0.0                   lazyeval_0.2.1             
## [15] colorspace_1.3-2            withr_2.1.2                
## [17] mnormt_1.5-5                tidyselect_0.2.4           
## [19] prettyunits_1.0.2           extrafontdb_1.0            
## [21] bit_1.1-14                  compiler_3.5.0             
## [23] cli_1.0.0                   rvest_0.3.2                
## [25] xml2_1.2.0                  DelayedArray_0.6.4         
## [27] rtracklayer_1.40.4          scales_1.0.0               
## [29] pbivnorm_0.6.0              digest_0.6.15              
## [31] Rsamtools_1.32.2            rmarkdown_1.10             
## [33] XVector_0.20.0              pkgconfig_2.0.1            
## [35] htmltools_0.3.6             extrafont_0.17             
## [37] rlang_0.2.1                 readxl_1.1.0               
## [39] rstudioapi_0.7              RSQLite_2.1.1              
## [41] bindr_0.1.1                 jsonlite_1.5               
## [43] BiocParallel_1.14.2         RCurl_1.95-4.11            
## [45] GenomeInfoDbData_1.1.0      munsell_0.5.0              
## [47] stringi_1.1.7               yaml_2.2.0                 
## [49] SummarizedExperiment_1.10.1 zlibbioc_1.26.0            
## [51] plyr_1.8.4                  grid_3.5.0                 
## [53] blob_1.1.1                  crayon_1.3.4               
## [55] lattice_0.20-35             Biostrings_2.48.0          
## [57] haven_1.1.2                 hms_0.4.2                  
## [59] knitr_1.20                  pillar_1.3.0               
## [61] codetools_0.2-15            biomaRt_2.36.1             
## [63] XML_3.98-1.15               glue_1.3.0                 
## [65] evaluate_0.11               modelr_0.1.2               
## [67] Rttf2pt1_1.3.7              cellranger_1.1.0           
## [69] gtable_0.2.0                papaja_0.1.0.9709          
## [71] assertthat_0.2.0            broom_0.5.0                
## [73] iterators_1.0.10            GenomicAlignments_1.16.0   
## [75] memoise_1.1.0               bindrcpp_0.2.2
\end{verbatim}

\hypertarget{app:covmat}{%
\section{Covariance matrices for the illustrative simulations}\label{app:covmat}}

\begin{table}[H]

\caption{\label{tab:cmatsup}The marginal covariance matrix for the data of the first illustrative simulation (suppression).}
\centering
\begin{tabular}{rrrr}
\toprule
$X$ & $M_1$ & $M_2$ & $Y$\\
\midrule
1.000 & -0.40 & 0.4 & -0.064\\
-0.400 & 1.00 & -0.6 & 0.256\\
0.400 & -0.60 & 1.0 & 0.000\\
-0.064 & 0.26 & 0.0 & 1.000\\
\bottomrule
\end{tabular}
\end{table}

\begin{table}[H]

\caption{\label{tab:cmatnoia}The marginal covariance matrix for the data of the second illustrative simulation.}
\centering
\resizebox{\linewidth}{!}{
\begin{tabular}{rrrrrrrrrrrrrrrrrr}
\toprule
$X$ & $M_{1}$ & $M_{2}$ & $M_{3}$ & $M_{4}$ & $M_{5}$ & $M_{6}$ & $M_{7}$ & $M_{8}$ & $M_{9}$ & $M_{10}$ & $M_{11}$ & $M_{12}$ & $M_{13}$ & $M_{14}$ & $M_{15}$ & $M_{16}$ & $Y$\\
\midrule
1.00 & 0.30 & -0.800 & -0.800 & 0.80 & -0.4000 & -0.400 & 0.400 & 0.400 & 0.400 & -0.400 & 0.400 & 0.400 & 0.4000 & -0.400 & -0.400 & 0.400 & 0.090\\
0.30 & 1.00 & -0.243 & -0.259 & 0.33 & -0.1682 & -0.137 & 0.147 & 0.181 & 0.177 & -0.244 & 0.124 & 0.142 & 0.1890 & -0.227 & -0.152 & 0.166 & 0.300\\
-0.80 & -0.24 & 1.000 & 0.650 & -0.64 & 0.3135 & 0.292 & -0.274 & -0.340 & -0.319 & 0.316 & -0.334 & -0.348 & -0.3931 & 0.351 & 0.315 & -0.353 & -0.073\\
-0.80 & -0.26 & 0.650 & 1.000 & -0.67 & 0.3729 & 0.306 & -0.309 & -0.283 & -0.319 & 0.316 & -0.293 & -0.345 & -0.2953 & 0.311 & 0.330 & -0.318 & -0.078\\
0.80 & 0.33 & -0.644 & -0.675 & 1.00 & -0.3311 & -0.329 & 0.339 & 0.314 & 0.328 & -0.332 & 0.259 & 0.342 & 0.3596 & -0.345 & -0.294 & 0.310 & 0.100\\
\addlinespace
-0.40 & -0.17 & 0.314 & 0.373 & -0.33 & 1.0000 & 0.249 & -0.159 & -0.196 & -0.212 & 0.220 & -0.155 & -0.218 & -0.0024 & 0.133 & 0.244 & -0.165 & -0.050\\
-0.40 & -0.14 & 0.292 & 0.306 & -0.33 & 0.2486 & 1.000 & -0.146 & -0.298 & -0.195 & 0.045 & -0.203 & -0.261 & -0.2198 & 0.222 & 0.016 & -0.169 & -0.041\\
0.40 & 0.15 & -0.274 & -0.309 & 0.34 & -0.1589 & -0.146 & 1.000 & 0.152 & 0.227 & -0.201 & 0.025 & 0.070 & 0.1157 & -0.150 & -0.150 & 0.170 & 0.044\\
0.40 & 0.18 & -0.340 & -0.283 & 0.31 & -0.1963 & -0.298 & 0.152 & 1.000 & 0.180 & -0.014 & 0.204 & 0.325 & 0.1817 & -0.156 & -0.146 & 0.317 & 0.054\\
0.40 & 0.18 & -0.319 & -0.319 & 0.33 & -0.2118 & -0.195 & 0.227 & 0.180 & 1.000 & -0.175 & -0.027 & 0.249 & 0.1384 & -0.226 & -0.132 & 0.064 & 0.053\\
\addlinespace
-0.40 & -0.24 & 0.316 & 0.316 & -0.33 & 0.2199 & 0.045 & -0.201 & -0.014 & -0.175 & 1.000 & -0.120 & -0.127 & -0.1484 & 0.091 & 0.175 & -0.124 & -0.073\\
0.40 & 0.12 & -0.334 & -0.293 & 0.26 & -0.1547 & -0.203 & 0.025 & 0.204 & -0.027 & -0.120 & 1.000 & 0.161 & 0.1668 & -0.257 & -0.167 & 0.242 & 0.037\\
0.40 & 0.14 & -0.348 & -0.345 & 0.34 & -0.2183 & -0.261 & 0.070 & 0.325 & 0.249 & -0.127 & 0.161 & 1.000 & 0.1782 & 0.025 & -0.222 & 0.179 & 0.043\\
0.40 & 0.19 & -0.393 & -0.295 & 0.36 & -0.0024 & -0.220 & 0.116 & 0.182 & 0.138 & -0.148 & 0.167 & 0.178 & 1.0000 & -0.295 & -0.036 & 0.210 & 0.057\\
-0.40 & -0.23 & 0.351 & 0.311 & -0.35 & 0.1333 & 0.222 & -0.150 & -0.156 & -0.226 & 0.091 & -0.257 & 0.025 & -0.2949 & 1.000 & 0.124 & -0.148 & -0.068\\
\addlinespace
-0.40 & -0.15 & 0.315 & 0.330 & -0.29 & 0.2437 & 0.016 & -0.150 & -0.146 & -0.132 & 0.175 & -0.167 & -0.222 & -0.0365 & 0.124 & 1.000 & -0.196 & -0.046\\
0.40 & 0.17 & -0.353 & -0.318 & 0.31 & -0.1654 & -0.169 & 0.170 & 0.317 & 0.064 & -0.124 & 0.242 & 0.179 & 0.2099 & -0.148 & -0.196 & 1.000 & 0.050\\
0.09 & 0.30 & -0.073 & -0.078 & 0.10 & -0.0505 & -0.041 & 0.044 & 0.054 & 0.053 & -0.073 & 0.037 & 0.043 & 0.0567 & -0.068 & -0.046 & 0.050 & 0.180\\
\bottomrule
\end{tabular}}
\end{table}

\begin{table}[H]

\caption{\label{tab:cmatnoib}The marginal covariance matrix for the data of the third illustrative simulation.}
\centering
\resizebox{\linewidth}{!}{
\begin{tabular}{rrrrrrrrrrrrrrrrrr}
\toprule
$X$ & $M_{1}$ & $M_{2}$ & $M_{3}$ & $M_{4}$ & $M_{5}$ & $M_{6}$ & $M_{7}$ & $M_{8}$ & $M_{9}$ & $M_{10}$ & $M_{11}$ & $M_{12}$ & $M_{13}$ & $M_{14}$ & $M_{15}$ & $M_{16}$ & $Y$\\
\midrule
1.00 & 0.3000 & 0.0000 & 0.0000 & 0.000 & 0.0000 & 0.000 & 0.0000 & 0.0000 & 0.0000 & 0.000 & 0.0000 & 0.000 & 0.0000 & 0.0000 & 0.0000 & 0.0000 & 0.09\\
0.30 & 1.0000 & -0.0096 & -0.0517 & 0.255 & -0.0574 & -0.020 & 0.0319 & 0.0728 & 0.0680 & -0.148 & 0.0043 & 0.027 & 0.0821 & -0.1275 & -0.0381 & 0.0551 & 0.85\\
0.00 & -0.0096 & 1.0000 & 0.0789 & -0.028 & -0.0215 & -0.091 & 0.1532 & -0.0647 & 0.0034 & -0.014 & -0.0479 & -0.091 & -0.2418 & 0.1037 & -0.0179 & -0.1095 & -1.03\\
0.00 & -0.0517 & 0.0789 & 1.0000 & -0.268 & 0.1749 & -0.047 & 0.0355 & 0.1234 & 0.0038 & -0.013 & 0.0889 & -0.084 & 0.0816 & -0.0292 & 0.0329 & 0.0070 & -1.04\\
0.00 & 0.2552 & -0.0278 & -0.2675 & 1.000 & -0.0368 & -0.029 & 0.0619 & -0.0183 & 0.0266 & -0.038 & -0.2009 & 0.073 & 0.1311 & -0.0840 & 0.0863 & -0.0343 & 1.17\\
\addlinespace
0.00 & -0.0574 & -0.0215 & 0.1749 & -0.037 & 1.0000 & 0.126 & 0.0016 & -0.0515 & -0.0733 & 0.085 & 0.0076 & -0.083 & 0.2233 & -0.0378 & 0.1186 & -0.0077 & -0.68\\
0.00 & -0.0204 & -0.0912 & -0.0473 & -0.029 & 0.1256 & 1.000 & 0.0196 & -0.1951 & -0.0490 & -0.163 & -0.0613 & -0.143 & -0.0847 & 0.0883 & -0.2037 & -0.0122 & -0.47\\
0.00 & 0.0319 & 0.1532 & 0.0355 & 0.062 & 0.0016 & 0.020 & 1.0000 & -0.0112 & 0.0949 & -0.058 & -0.1907 & -0.127 & -0.0627 & 0.0146 & 0.0145 & 0.0137 & 0.20\\
0.00 & 0.0728 & -0.0647 & 0.1234 & -0.018 & -0.0515 & -0.195 & -0.0112 & 1.0000 & 0.0281 & 0.207 & 0.0624 & 0.233 & 0.0307 & 0.0051 & 0.0191 & 0.2226 & 0.59\\
0.00 & 0.0680 & 0.0034 & 0.0038 & 0.027 & -0.0733 & -0.049 & 0.0949 & 0.0281 & 1.0000 & -0.021 & -0.2649 & 0.126 & -0.0306 & -0.0941 & 0.0397 & -0.1354 & 0.44\\
\addlinespace
0.00 & -0.1482 & -0.0140 & -0.0132 & -0.038 & 0.0849 & -0.163 & -0.0585 & 0.2070 & -0.0210 & 1.000 & 0.0572 & 0.046 & 0.0164 & -0.0971 & 0.0210 & 0.0509 & -0.27\\
0.00 & 0.0043 & -0.0479 & 0.0889 & -0.201 & 0.0076 & -0.061 & -0.1907 & 0.0624 & -0.2649 & 0.057 & 1.0000 & 0.002 & 0.0097 & -0.1377 & -0.0093 & 0.1163 & 0.16\\
0.00 & 0.0267 & -0.0913 & -0.0842 & 0.073 & -0.0826 & -0.143 & -0.1274 & 0.2333 & 0.1260 & 0.046 & 0.0020 & 1.000 & 0.0258 & 0.2620 & -0.0883 & 0.0270 & 0.72\\
0.00 & 0.0821 & -0.2418 & 0.0816 & 0.131 & 0.2233 & -0.085 & -0.0627 & 0.0307 & -0.0306 & 0.016 & 0.0097 & 0.026 & 1.0000 & -0.1913 & 0.1750 & 0.0707 & 0.62\\
0.00 & -0.1275 & 0.1037 & -0.0292 & -0.084 & -0.0378 & 0.088 & 0.0146 & 0.0051 & -0.0941 & -0.097 & -0.1377 & 0.262 & -0.1913 & 1.0000 & -0.0505 & 0.0168 & -0.58\\
\addlinespace
0.00 & -0.0381 & -0.0179 & 0.0329 & 0.086 & 0.1186 & -0.204 & 0.0145 & 0.0191 & 0.0397 & 0.021 & -0.0093 & -0.088 & 0.1750 & -0.0505 & 1.0000 & -0.0510 & -0.27\\
0.00 & 0.0551 & -0.1095 & 0.0070 & -0.034 & -0.0077 & -0.012 & 0.0137 & 0.2226 & -0.1354 & 0.051 & 0.1163 & 0.027 & 0.0707 & 0.0168 & -0.0510 & 1.0000 & 0.60\\
0.09 & 0.8462 & -1.0314 & -1.0374 & 1.169 & -0.6790 & -0.468 & 0.1981 & 0.5929 & 0.4422 & -0.272 & 0.1592 & 0.724 & 0.6195 & -0.5760 & -0.2686 & 0.5984 & 10.17\\
\bottomrule
\end{tabular}}
\end{table}

\newpage

\end{document}